\documentclass[prd,aps,a4paper,nofootinbib,twocolumn]{revtex4}

\newif\ifusesec
\usesectrue  
   
\usepackage{graphicx} 
\usepackage{mathrsfs}
\usepackage{amsmath,amsfonts,amssymb}
\usepackage{multirow}

\newcommand{\beq}{\begin{equation}}
\newcommand{\eeq}{\end{equation}}

\begin{document}

\title{New gravitational self-force analytical results for eccentric equatorial orbits around a Kerr black hole: gyroscope precession}

\author{Donato \surname{Bini}}
\author{Andrea \surname{Geralico}}
\affiliation{$^1$Istituto per le Applicazioni del Calcolo ``M. Picone,'' CNR, I-00185 Rome, Italy}

\date{\today}

\begin{abstract}
We analytically compute the gravitational self-force correction to the gyroscope precession along slightly eccentric equatorial orbits in the Kerr spacetime, generalizing previous results for the Schwarzschild spacetime. 
Our results are accurate through the 9.5 post-Newtonian order and to second order in both eccentricity and rotation parameter.
We also provide a post-Newtonian check of our results based on the currently known Hamiltonian for spinning binaries. 
\end{abstract}

\maketitle

\section{Introduction}

The last few years have witnessed the beginning of the era of gravitational-wave astronomy, after the discovery of the first signals by LIGO \cite{Abbott:2016blz,Abbott:2016nmj,Abbott:2017oio,TheLIGOScientific:2017qsa,Abbott:2017vtc,Abbott:2017gyy} associated with either binary black hole or neutron star mergers.
The number of such events is expected to rapidly increase in the near future thanks to the improved sensitivity of Advanced LIGO \cite{aligo} and to the contribution of the space-based interferometer eLISA \cite{elisa}, which is designed to detect a wide range of low-frequency gravitational wave sources, including extreme mass ratio
inspirals (EMRIs). The latter are binary systems in which one body is much more massive than the other, so that the dynamics is well described in the framework of gravitational self-force (GSF) theory by using standard first-order perturbation methods (see, e.g., Ref. \cite{Barack:2018yvs} for a recent review).
Conservative effects are encoded in gauge-invariant quantities, which are insensitive of the particular method used to perform the calculation and of the chosen technique to regularize and fully reconstruct the metric perturbation.
These invariant thus provide useful information which can be used to compare results from other approaches, like Post-Newtonian (PN) theory and numerical relativity (NR) simulations, as well as to calibrate and enhance the Effective-One-Body (EOB) model \cite{Buonanno:1998gg,Buonanno:2000ef,Damour:2001tu}.

Spin couplings are expected to significantly affect the two-body dynamics, thereby playing an important role in the gravitational wave detection and parameter estimation (see, e.g., Ref. \cite{Blanchet:2013haa} and references therein).
Spin-orbital, i.e., linear-in-spin, and spin-spin, i.e., quadratic-in-spin, effects have been accounted at the lowest PN levels by standard Hamiltonian methods \cite{Damour:2007nc,Barausse:2009aa,Vines:2016unv} and effective field theory (EFT) techniques \cite{Levi:2015msa,Levi:2016ofk}.
The first high-PN calculations within the GSF approach of the spin-orbit precession of a spinning compact body on a circular orbit around a Schwarzschild black hole have been done in Refs. \cite{Dolan:2013roa,Bini:2014ica,Bini:2015mza}.
These results have been extended to eccentric orbits in Refs. \cite{Kavanagh:2017wot,Bini:2018aps} by using the methodology introduced in Ref. \cite{Akcay:2016dku}, soon after generalized to the Kerr case in Ref. \cite{Akcay:2017azq}.

We compute here the GSF correction to the spin-precession invariant for slightly eccentric equatorial orbits in the Kerr spacetime through the 9.5 PN order and to second order both in the eccentricity and spin parameter.
The spin-dependent part mixing eccentricity and spin effects is completely new.
We also improve to the 9.5 PN level the current knowledge of the spin-precession invariant for eccentric orbits in the non-rotating case (9 PN, Ref. \cite{Bini:2018aps}) and for circular orbits in the same Kerr case (8 PN, Ref. \cite{Bini:2018ylh}) up the second order in the spin parameter.
Furthermore, the circular orbit limit of the present result gives the self-force correction to the periastron advance around a Kerr black hole, which has been presented elsewhere \cite{Bini2019:deltakcirc}.
Finally, as an independent check, we calculate the same invariant by using the current knowledge of the Arnowitt-Deser-Misner (ADM) Hamiltonian for two point masses with aligned spins \cite{Schafer:2018kuf}. 

We will denote by $m_1$ and $m_2$ and by $S_1$ and $S_2$ the masses and spins of the two bodies, respectively, with the convention that $m_1\le m_2$.
We also define the total mass of the system $M=m_1+m_2$, the mass ratios 
\beq
q=\frac{m_1}{m_2}\,,\qquad
\mu =\frac{m_1m_2}{M}\,,\qquad
\nu =\frac{\mu}{M}=\frac{q}{(1+q)^2}
\,,
\eeq
and the dimensionless mass difference 
\beq
\frac{m_2-m_1}{M}=\Delta=\sqrt{1-4\nu}\,,
\eeq
as well as the dimensionless spin variables $\chi_{1,2}\equiv S_{1,2}/m_{1,2}^2$ associated with each body, as usual.
GSF results are obtained in the limit of small mass-ratio ($m_1\ll m_2$, implying $q\sim\nu\ll 1$) and small spin ($|S_1|/(cGm_1^2)\ll 1$) of the perturbing body. 
The metric signature is chosen to be $+2$ and units are such that $c=G=1$ unless differently specified.
Greek indices run from 0 to 3, whereas Latin ones from 1 to 3.

\section{Gyroscope precession in the background Kerr spacetime}

The background Kerr metric with parameters $m_2$ and $a_2=a$ (with $\hat a =a/m_2$ dimensionless) written in Boyer-Lindquist coordinates $(t,r,\theta,\phi)$ reads
\begin{eqnarray}
\label{kerrmet}
d{\bar s}^2&=&{\bar g}_{\alpha\beta}dx^\alpha dx^\beta\nonumber\\
&=&-\left(1-\frac{2m_2r}{\Sigma}  \right) dt^2-\frac{4am_2r \sin^2\theta}{\Sigma}dtd\phi\nonumber\\
&+& \frac{\Sigma}{\Delta}dr^2+\Sigma d\theta^2\nonumber\\
&+&  \left( r^2+a^2+\frac{2m_2ra^2\sin^2\theta}{\Sigma} \right)\sin^2\theta d\phi^2\,,
\end{eqnarray}
where   
\beq
\Delta= r^2+a^2-2m_2r\,,\qquad 
\Sigma=r^2+a^2\cos^2\theta\,.
\eeq
A test gyroscope moving along an eccentric geodesic orbit on the equatorial plane ($\theta=\pi/2$) has four velocity 
\begin{eqnarray}
\label{barudef}
\bar u&=&\bar u^\alpha\partial_\alpha
=\frac{1}{r^2} \left(a x+\frac{r^2+a^2}{\Delta} \bar P\right)\partial_t + \dot r\partial_r \nonumber\\
&&
+\frac{1}{r^2}\left(x+\frac{a}{\Delta} \bar P\right)\partial_\phi
\,,
\end{eqnarray}
where $\bar P=\bar E r^2 -a x$, with $x=\bar L-a\bar E$, and $\dot r \equiv \bar u^r$ is such that
\beq
\label{eq_r}
\dot r^2=\left(\frac{dr}{d\bar\tau}\right)^2
=\frac{1}{r^4}\left[\bar P^2-\Delta(r^2+x^2)\right] 
\,.
\eeq
Here $\bar E=-\bar u_t$ and $\bar L= \bar u_\phi$ denote the conserved energy and angular momentum per unit mass of the particle, respectively, so that $\bar E$ and $\bar L/m_2$ are dimensionless together with their combination $\hat x=x/m_2$.
The orbit can be parametrized either by the proper time $\bar \tau$ or by the relativistic anomaly $\chi\in[0,2\pi]$, such that 
\beq
r=\frac{m_2 p}{1+e\cos \chi}\,,
\eeq
which are related by 
\beq
\label{dtaudchi}
m_2 \frac{d\chi}{d\bar\tau} =  u_p^{3/2}(1+e\cos \chi )^2
[1+u_p^2\, \hat x{}^2 ( e^2-2 e\cos\chi-3)]^{1/2}\,.
\eeq
The (dimensionless) background orbital parameters, semi-latus rectum $p$ (with reciprocal $u_p=1/p$) and eccentricity $e$, are defined by writing the minimum (pericenter,
$r_{\rm peri}$) and maximum (apocenter, $r_{\rm apo}$) values of the radial coordinate along the orbit as 
\beq
r_{\rm peri}=\frac{m_2 p}{1+e}\,,\qquad r_{\rm apo}=\frac{m_2 p}{1-e}\,.
\eeq
The two conditions
\beq
\label{drextrema}
\left(\frac{dr}{d\bar\tau}\right)\bigg|_{r_{\rm peri}}=0=\left(\frac{dr}{d\bar\tau}\right)\bigg|_{r_{\rm apo}}\,,
\eeq
can be imposed on Eq.~\eqref{eq_r} to solve them for $\bar E=\bar E(p,e)$ and $\bar L=\bar L(p,e)$.
Their explicit expressions in terms of ($u_p, e, \hat a$) for prograde orbits are given by 
\begin{widetext}
\begin{eqnarray}
\bar E&=&\frac{1-2u_p+{\hat a}u_p^{3/2}}{\sqrt{1-3u_p+2{\hat a}u_p^{3/2}}}\left\{
1-
\left[
\frac12
-\frac{2{\hat a}u_p^{5/2}}{(1-2u_p+{\hat a}^2u_p^2)(1-2u_p)}
+\frac{1-4u_p}{2(1-3u_p+2{\hat a}u_p^{3/2})}\right.\right.\nonumber\\
&&\left.\left.
-\frac{1-4u_p+2u_p^2}{(1-2{\hat a}u_p^{3/2}+{\hat a}^2u_p^2)(1-2u_p)}
\right]e^2
\right\}
+O(e^4)
\,,\nonumber\\
\frac{\bar L}{m_2}&=&\frac{1-2{\hat a}u_p^{3/2}+{\hat a}^2u_p^2}{\sqrt{u_p(1-3u_p+2{\hat a}u_p^{3/2})}}\left\{
1-
\left[
\frac12
+\frac{{\hat a}u_p^{1/2}(1+u_p)}{1-2u_p+{\hat a}^2u_p^2}
+\frac{1-4u_p}{2(1-3u_p+2{\hat a}u_p^{3/2})}
-\frac{1+{\hat a}u_p^{1/2}(1-u_p)}{1-2{\hat a}u_p^{3/2}+{\hat a}^2u_p^2}
\right]e^2
\right\}\nonumber\\
&&
+O(e^4)
\,,\nonumber\\
\end{eqnarray}
respectively, to the second order in eccentricity.

The motion is then governed by the following equations \cite{Glampedakis:2002ya,Bini:2016iym}
\begin{eqnarray}
\frac{dt}{d\chi}&=&\frac{m_2}{u_p^{3/2}}\frac{E +E\hat a^2 u_p^2 (1+e\cos\chi)^2   -2 \hat a u_p^3\hat x (1+e\cos\chi)^3}{(1+e\cos \chi )^2[1+u_p^2 \,\hat x{}^2 (e^2-2 e\cos \chi -3) ]^{1/2}
[1-2 u_p(1+ e\cos \chi) +a^2 u_p^2(1+ e\cos \chi)^2  ]}
\,,\nonumber\\
\frac{d\phi}{d\chi}&=& u_p^{1/2}\frac{ \hat x + \hat a E - 2 u_p \hat x (1+  e\cos \chi) }{[1+u_p^2 \,\hat x{}^2 (e^2-2 e\cos \chi -3) ]^{1/2}
[1-2 u_p(1+ e\cos \chi) +a^2 u_p^2(1+ e\cos \chi)^2  ]}
\,.
\end{eqnarray}
\end{widetext}
Integrating over a full radial orbit from periastron to periastron gives the coordinate time radial period $\bar T_{r}=\oint dt= \oint d\chi (dt/d\chi)$ and the accumulated azimuthal angle $\bar\Phi=\oint d\phi = \oint d\chi (d\phi/d\chi)$, with associated frequencies  $\bar\Omega_{r}=2\pi/\bar T_{r}$ and $\bar\Omega_{\phi}=\bar\Phi/\bar T_{r}$.

\subsection{Marck's \lq\lq intermediate'' frame and gyroscope precession}

Using the Killing-Yano tensor  Marck defined a parallely propagated frame along a general geodesic in the Kerr spacetime \cite{Marck}. 
Marck's geometric construction uses, as an \lq\lq intermediate" frame, a convenient (degenerate) Frenet-Serret frame adapted to $\bar u$, which in the case of equatorial timelike geodesics reads
\begin{eqnarray}
\bar e_1 &=& \frac{r}{ (r^2+x^2)^{1/2}} \left[\frac{ \dot r (r^2+a^2)}{ \Delta  }\left(\partial_t   +\frac{a}{r^2+a^2}\partial_\phi\right)\right.\nonumber\\
&&\left.
+\frac{\bar P}{r^2 }  \partial_r \right]
\,,\nonumber\\
\bar e_2 &=& \frac{1}{r}\partial_\theta
\,, \nonumber\\
\bar e_3 &=&  \left(\frac{x (r^2+a^2)\bar P}{ (r^2+x^2)^{1/2} \Delta  r^2}+\frac{a (r^2+x^2)^{1/2}}{r^2} \right)\partial_t\nonumber\\
&&
 + \frac{x\dot r}{(r^2+x^2)^{1/2}} \partial_r \nonumber\\
&&
+\left( \frac{ax\bar P}{(r^2+x^2)^{1/2}r^2 \Delta }+\frac{(r^2+x^2)^{1/2}}{r^2} \right) \partial_\phi\,,
\end{eqnarray}
whose transport properties are
\beq
\nabla_{\bar u} \bar e_1 =\bar \omega \bar e_3\,,\qquad \nabla_{\bar u} \bar e_3 =-\bar \omega \bar e_1\,,
\eeq
with 
\beq
\bar \omega=\frac{\bar E x+ a}{r^2+x^2}\,,
\eeq
whereas $\nabla_{\bar u} \bar e_2=0$, since $\bar e_2$ is aligned with the $\theta$-direction.
The total spin precession angle accumulated over a radial period is then 
\beq
\bar \Psi=\int_0^{\bar{\mathcal T}_r} \bar \omega d\bar \tau
=\int_0^{2\pi} \bar \omega \frac{d\bar \tau}{d\chi}d\chi\,,
\eeq
$\bar{\mathcal T}_r=\oint d\bar\tau$ denoting the proper-time period.
In order to remove the rotation of the Boyer-Lindquist spherical-like coordinate frame in the azimuthal direction, corresponding to comparing the spin direction with a \lq\lq fixed'' asymptotic Cartesian-like frame, one must subtract $\bar \Phi$ from 	$\bar \Psi$.
The net precession angle of the test gyroscope dragged along $\bar u$ is then conveniently measured by the quantity
\beq
\bar \psi=1-\frac{\bar \Psi}{\bar \Phi}\,,
\eeq
which reads
\begin{widetext}
\begin{eqnarray}
\bar \psi&=&1-\sqrt{1-3u_p+2{\hat a}u_p^{3/2}}
+\frac{3u_p^2(1-{\hat a}u_p^{1/2})^2}{2(1-6u_p+8{\hat a}u_p^{3/2}-3{\hat a}^2u_p^2)^2\sqrt{1-3u_p+2{\hat a}u_p^{3/2}}(1-2u_p+{\hat a}^2u_p^2)}\nonumber\\
&&\times
\left[(1-6u_p)(1-4u_p)+2(5-22u_p)u_p^{3/2}{\hat a}+10{\hat a}^2u_p^3-2(1-15u_p)u_p^{5/2}{\hat a}^3-25{\hat a}^4u_p^4+6{\hat a}^5u_p^{9/2}
\right]\,e^2
+O(e^4)
\,,\nonumber\\
\end{eqnarray}
\end{widetext}
to the second order in the eccentricity parameter.

\section{Spin precession in the perturbed spacetime}

In this section we recall the basic theory underlying the derivation of the spin precession invariant in the perturbed spacetime and its first-order SF correction, following Refs. \cite{Akcay:2016dku,Akcay:2017azq}. 
The gyroscope carrying a small mass $m_1$ and a small spin $S_1$ (so that $q = \frac{m_1}{m_2} \ll 1$ and $|S_1|/(cGm_1^2)\ll 1$) can be considered as following an eccentric geodesic orbit in a (regularized) perturbed spacetime $g^{\rm R}_{\alpha\beta}$, through order $O(q)$, while its associated spin vector is parallely-transported in $g^{\rm R}_{\alpha\beta}$, to linear order in spin. 
The regularized perturbed metric is decomposed as
\beq
\label{pertmet}
g^{\rm R}_{\alpha\beta}=\bar g_{\alpha\beta}+h^{\rm R}_{\alpha\beta} + O(q^2)\,,
\eeq
where $\bar g_{\alpha\beta}$ is the background spacetime \eqref{kerrmet} and $h^{\rm R}_{\alpha\beta}=O(q)$  is the first-order SF metric perturbation. Henceforth, we shall omit the superscript R.
The spin precession invariant
\beq
\label{psi_def}
\psi(m_2\Omega_r, m_2\Omega_\phi; q)=1-\frac{\Psi}{\Phi} 
\,,
\eeq
is assumed to be a function of the the radial and (averaged) azimuthal angular frequencies $\Omega_r=2\pi/T_r$ and $\Omega_{\phi}=\Phi/T_r$, for any value of the mass ratio.
Furthermore, the geodesics in both background and perturbed spacetimes are assumed to have the same orbital parameters ($p, e$), so that any comparison between perturbed and unperturbed quantities is done at the same coordinate radius $r$ (or the same anomaly $\chi$), though not the same $t$ and $\phi$ coordinates. Any such difference is not gauge-invariant, in general. Gauge invariance is ensured by further assuming that the background and perturbed orbits both have the same orbital frequencies (or equivalently the same radial and azimuthal periods).
The first-order SF correction to the spin precession invariant is then defined as
\begin{eqnarray}
\label{Deltapsidef}
\Delta \psi&=& \frac1q \left[\psi(m_2\Omega_r,m_2\Omega_{\phi};q)-\psi(m_2\Omega_r,m_2\Omega_{\phi};0) \right]\nonumber\\
&=&-\frac{\Delta\Psi}{\Phi}\,,
\end{eqnarray}
where
\beq
\label{DeltaPsidef}
\Delta \Psi=\delta \Psi-\frac{\partial \bar \Psi}{\partial \bar \Omega_r}\delta \Omega_r -\frac{\partial \bar \Psi}{\partial \bar \Omega_\phi}\delta \Omega_\phi\,,
\eeq
the operator $\delta$ denoting the $O(q)$ difference between a quantity on the perturbed geodesic and the same quantity on the background one with the same $(p,e,\chi)$, but which does not keep fixed the values of the two frequencies.
After the computation of the function $\Delta \psi(\Omega_r,\Omega_{\phi})$, one can reexpress it as a function of the inverse semi-latus rectum $u_p$, and eccentricity $e$, of the unperturbed orbit.

\subsection{Bound timelike geodesics}

Bound timelike geodesics in the equatorial plane of the perturbed spacetime \eqref{pertmet} have $4$-velocity
\beq
u=u^\alpha\partial_\alpha=(\bar u^\alpha+\delta u^\alpha)\partial_\alpha\,, 
\eeq
with $\delta u^\alpha =O(h)$, and $u^\theta=0=\bar u^\theta$, so that $\delta u^\theta=0$.
Let us introduce the first order quantities $\delta E$ and $\delta L$ such that the four velocity components $u^\alpha$ can be written exactly in the same form as those of the background \eqref{barudef} with the replacement $\bar E\to \bar E+\delta E$ and $\bar L\to \bar L+\delta L$, implying that 
\begin{eqnarray}
 \delta u^t 
&=& \left[\frac{(r^2+a^2)^2}{\Delta}-a^2  \right]\frac{ \delta E}{r^2}+\left[1-\frac{r^2+a^2}{\Delta}  \right]\frac{a}{r^2} \delta L
\,,\nonumber\\
 \delta u^\phi  
&=& \left[ \frac{r^2+a^2}{\Delta}-1 \right]\frac{a}{r^2} \delta E+\left[1-\frac{a^2}{\Delta}  \right]\frac{ \delta L}{r^2}\,,
\end{eqnarray} 
which can be in turn inverted to yield $\delta E=-\bar g_{t\alpha} \delta u^\alpha$ and $\delta L = \bar g_{\phi\alpha} \delta u^\alpha$.
The correction $\delta u^r$ to the radial component of the four velocity directly follows from the normalization condition of $u$ ($u\cdot u=-1$) with respect to the perturbed metric, which reads
\beq
\bar g_{rr} \bar u^r \delta u^r = \bar u^t \delta E- \bar u^\phi \delta L-\frac12 h_{00}\,,
\eeq
where $h_{00}=h_{\alpha\beta}\bar u^\alpha \bar u^\beta$.
Equivalently, one can normalize $u$ with respect to the background metric as in Barack and Sago \cite{Barack:2011ed} (a hat denoting the corresponding quantities), implying
\beq
\delta u^\alpha =\hat \delta u^\alpha  +\frac12 h_{00}\bar u^\alpha\,,
\eeq
leading to the relations
\begin{eqnarray}
\label{relwithBS}
\hat\delta E&=&\delta E-\frac12\bar E h_{00}\,, \nonumber\\
\hat\delta u^r&=&\delta u^r-\frac12\bar u^r h_{00}\,, \nonumber\\
\hat\delta L&=&\delta L-\frac12\bar L h_{00}\,,
\end{eqnarray}
with
\beq
\label{delta_ur_BS}
\bar g_{rr} \bar u^r \hat\delta u^r = \bar u^t \hat\delta E- \bar u^\phi \hat\delta L\,.
\eeq

The geodesic equations 
\beq
\frac{du_\alpha}{d\tau}-\frac12 (\bar g_{\lambda\mu,\alpha}+h_{\lambda\mu,\alpha})u^\lambda u^\mu=0\,,
\eeq  
with
\beq
u_\alpha =\bar u_\alpha +h_{0\alpha}+\bar g_{\alpha\beta}\delta u^\beta\,,
\eeq
determine the evolution of $\delta u_t$ and $\delta u_\phi$, or equivalently of the perturbations in
energy $\hat \delta E$ and angular momentum $\hat \delta L$ by
\begin{eqnarray}
\label{eqdeltaEL_BS}
\frac{d}{d\tau}\hat \delta E = -F_t \,,\qquad 
\frac{d}{d\tau}\hat \delta L = F_\phi\,,
\end{eqnarray}
where the functions $F_t$ and $F_\phi$ are the covariant $t$ and $\phi$ components of the self force
\begin{eqnarray}
F^\mu&=&-\frac12(\bar g^{\mu\nu}+\bar u^\mu\bar u^\nu)\bar u^\lambda\bar u^\rho(2h_{\nu\lambda;\rho}-h_{\lambda\rho;\nu})\nonumber\\
&\equiv&-\frac12P(\bar u)^{\mu\nu}\bar u^\lambda\bar u^\rho h_{\{\nu\lambda;\rho\}_-}\,,
\end{eqnarray}
the anticyclic permutation notation $A_{\{abc\}_-}=A_{abc}-A_{bca}+A_{cab}$ having been introduced.
Here we are interested in conservative effects only, i.e., we assume that $F^\alpha=F^\alpha_{\rm cons}$ results in a periodic function of $\chi$.
Eqs. \eqref{eqdeltaEL_BS} can then be formally integrated as
\begin{eqnarray}
\hat \delta E(\chi) &=& -\int_0^\chi F_t^{\rm cons}(\chi) \frac{d\tau}{d\chi}d\chi+\hat \delta E(0)
\nonumber\\
&\equiv& {\mathcal E}(\chi)+\hat \delta E(0)
\,,\nonumber\\
\hat \delta L(\chi) &=& \int_0^\chi F_\phi^{\rm cons}(\chi) \frac{d\tau}{d\chi}d\chi+\hat \delta L(0)
\nonumber\\
&\equiv& {\mathcal L}(\chi)+\hat \delta L(0)
\,,
\end{eqnarray}
 where the conservative SF components are defined by $F_t^{\rm cons}=[F_t(\chi)-F_t(-\chi)]/2$ and $F_\phi^{\rm cons}=[F_\phi(\chi)-F_\phi(-\chi)]/2$.
The integration constants $\hat \delta E(0)$ and $\hat \delta L(0)$ are computed by imposing the vanishing of $\hat\delta u^r$ both at the periastron ($\chi=0$) and the apoastron  ($\chi=\pi$), i.e.,
\begin{eqnarray}
\label{deltaELperiapo}
 0&=&\bar u^t(0)\hat \delta E(0)-\bar u^\phi(0) \hat \delta L(0)
 \,,\nonumber\\
 0&=&\bar u^t(\pi)\hat \delta E(\pi)- \bar u^\phi(\pi)\hat \delta L(\pi)\,, 
\end{eqnarray}
form Eq. \eqref{delta_ur_BS}, leading to
\begin{eqnarray}
\hat \delta E(0) 
&=& -\bar u^\phi(0)\frac{[-\bar u^t (\pi) {\mathcal E}(\pi)+\bar u^\phi (\pi) {\mathcal L}(\pi)]}{S(0,\pi)}\nonumber\\
\hat \delta L(0) 
&=&  -\bar u^t(0)\frac{[-\bar u^t (\pi) {\mathcal E}(\pi)+\bar u^\phi (\pi) {\mathcal L}(\pi)]}{S(0,\pi)}\,,
\end{eqnarray}
where $S(0,\pi)=\bar u^t(0)\bar u^\phi(\pi)-\bar u^t(\pi)\bar u^\phi(0)$.

\subsection{GSF corrections to the spin precession invariant}

The spin precession has been calculated in Ref. \cite{Akcay:2017azq} with respect to a suitably defined perturbed Marck-type frame $\{u,e_a\}$ adapted to $u$, with $e_a^\alpha=\bar e_a^\alpha +\delta e_a^\alpha$, with with $\delta e_a^\alpha =O(h)$.
The first-order SF correction $\Delta \psi$ to the spin precession invariant \eqref{Deltapsidef} is expressed in terms of the corresponding correction $\Delta \Psi$ to the amount of precession angle accumulated by the spin vector over one radial period defined by Eq. \eqref{DeltaPsidef}, where 
\beq
\delta \Psi=\int_0^{2\pi}  \left(\frac{\delta \omega}{\bar \omega}-\frac{\delta u^r}{\bar u^r}\right) \bar \omega \frac{d \bar \tau}{d\chi}  d\chi\,,
\eeq
whereas the SF corrections to the frequencies are given by
\beq
\label{deltaOmegas}
\delta \Omega_r=-\bar \Omega_r\frac{\delta T_r}{\bar T_r}\,,\qquad
\delta \Omega_\phi=-\bar \Omega_\phi\left(-\frac{\delta \Phi}{\bar \Phi}+\frac{\delta T_r}{\bar T_r}\right)\,,
\eeq 
with
\begin{eqnarray}
\label{deltaTePhi}
\delta T_r&=&\int_0^{2\pi}  \left(\frac{\delta u^t}{\bar u^t}-\frac{\delta u^r}{\bar u^r}\right) \bar u^t \frac{d \bar \tau}{d\chi}  d\chi
\,,\nonumber\\
\delta \Phi&=&\int_0^{2\pi}  \left(\frac{\delta u^\phi}{\bar u^\phi}-\frac{\delta u^r}{\bar u^r}\right) \bar u^\phi \frac{d \bar \tau}{d\chi}  d\chi\,.
\end{eqnarray}
The quantity $\delta\omega$ is defined in Eq. (3.20) of Ref. \cite{Akcay:2017azq}.
It can be conveniently rewritten as 
\beq
\hat \delta \omega = \delta \Gamma_{[31]0}+c_{01}\bar {\mathcal R}_{11,3}+c_{03}\bar {\mathcal R}_{13,3}\,,
\eeq
where $\hat\delta \omega=\delta \omega-\frac12\bar \omega h_{00}$,
\begin{eqnarray}
c_{01}&=& -\frac{ \hat \delta u^r \bar e_3^\phi-\bar e_3^r\hat \delta u^\phi }{ \bar e_1^\phi \bar e_3^r-\bar e_3^\phi \bar e_1^r } \,,\nonumber\\
c_{03}&=&\frac{ -\hat \delta u^\phi \bar e_1^r+\bar e_1^\phi  \hat \delta u^r }{ \bar e_1^\phi \bar e_3^r-\bar e_3^\phi \bar e_1^r} \,,
\end{eqnarray}
and 
\begin{eqnarray}
\bar {\mathcal R}_{11,3}&=&
\frac{x\sqrt{r^2+x^2}}{r\bar u_r}\left(\frac{M}{r^3}-\bar\omega^2\right)\,,\nonumber\\
\bar {\mathcal R}_{13,3}&=&
\frac{E+\bar\omega x}{\sqrt{r^2+x^2}}\,,
\end{eqnarray}
are the Ricci rotation coefficients of the background frame
\beq
\bar {\mathcal R}_{\beta\alpha,\sigma}=\bar e_\sigma \cdot_{\bar g} \nabla_{\bar e_\alpha}  \bar e_{\beta}\,.
\eeq
Finally, the quantity $\delta \Gamma_{[31]0}$ is explicitly given in Appendix B of Ref. \cite{Akcay:2017azq} in terms of the components of the metric perturbation and their first derivatives.

\section{Self-force calculation}

The procedure for obtaining the first order metric perturbations of a Kerr spacetime by using the Teukolsky formalism in a radiation gauge is well established in the literature (see, e.g., Refs. \cite{Keidl:2010pm,Shah:2012gu}).
This method has been already applied to the computation of the corrections to the gyroscope precession along eccentric orbits in a Schwarzschild spacetime in Refs. \cite{Kavanagh:2017wot,Bini:2018aps} and for circular orbits in the same Kerr case in Ref. \cite{Bini:2018ylh}.
Therefore, we refer to these works for a detailed account of all the intermediate steps, including the subtleties concerning the regularization technique (see Section IIIE of Ref. \cite{Kavanagh:2017wot} and Section IIIB of Ref. \cite{Bini:2018ylh}) as well as the completion of the metric perturbation (see Section IIIC of Ref. \cite{Bini:2018ylh}).
We provide below only the relevant information about the nonradiative multipoles and the regularization parameter used in our analysis.

The contribution of the lowest modes $l=0,1$ in the spacetime region inside (left, $-$) and outside (right, $+$) the particle's location turns out to be
\begin{widetext}
\begin{eqnarray}
\Delta\psi^-_{l=0,1}&=&
-\frac{(-4+42u_p-121u_p^2+98u_p^3)}{(-1+3u_p)(4-39u_p+86u_p^2)}u_p\nonumber\\
&&
+\frac{(24-558u_p+5281u_p^2-26410u_p^3+75061u_p^4-116396u_p^5+76996u_p^6)}{(-1+3u_p)^2(4-39u_p+86u_p^2)^2}u_p^{3/2}\hat a\nonumber\\
&&
+\frac{u_p^2}{(-1+3u_p)^3(4-39u_p+86u_p^2)^3}
(-32+1296u_p-22098u_p^2+208882u_p^3-1208315u_p^4+4451526u_p^5\nonumber\\
&&-10533213u_p^6+15691174u_p^7-13712636u_p^8+5540680u_p^9){\hat a}^2\nonumber\\
&&
+\bigg\{
\frac12\frac{(-40+1088u_p-12307u_p^2+75418u_p^3-273210u_p^4+594423u_p^5-732436u_p^6+400092u_p^7)}{(-1+3u_p)^2(4-39u_p+86u_p^2)^2(-1+2u_p)(-1+6u_p)}u_p^2\nonumber\\
&&
-\frac14\frac{u_p^{5/2}}{(-1+3u_p)^3(4-39u_p+86u_p^2)^3(-1+2u_p)(-1+6u_p)^2}
(-672+29032u_p-564560u_p^2\nonumber\\
&&
+6550123u_p^3-50644291u_p^4+275566031u_p^5-1082164705u_p^6+3075978930u_p^7\nonumber\\
&&
-6199296104u_p^8+8404397408u_p^9-6854125200u_p^{10}+2528579232u_p^{11})\hat a\nonumber\\
&&
-\frac14\frac{u_p^3}{(-1+3u_p)^4(4-39u_p+86u_p^2)^4(-1+2u_p)^2(-1+6u_p)^3}(-384-128u_p+703000u_p^2\nonumber\\
&&
-24934708u_p^3+457010141u_p^4-5436136756u_p^5+45820221810u_p^6-286235945992u_p^7\nonumber\\
&&+1358692923261u_p^8-4964206286808u_p^9+14012232755836u_p^{10}-30394437681256u_p^{11}\nonumber\\
&&+49854730541696u_p^{12}-59906488552896u_p^{13}+49754497440960u_p^{14}\nonumber\\
&&-25484985853056u_p^{15}+6047210836224u_p^{16}){\hat a}^2
\bigg\}e^2
+O({\hat a}^3,e^4)
\,,
\end{eqnarray}%
and
\begin{eqnarray}
\Delta\psi^+_{l=0,1}&=&
-\frac{(-4+36u_p-75u_p^2+14u_p^3)}{(-1+3u_p)(4-39u_p+86u_p^2)}u_p\nonumber\\
&&
+\frac{(24-486u_p+3959u_p^2-16766u_p^3+40403u_p^4-55220u_p^5+34588u_p^6)}{(-1+3u_p)^2(4-39u_p+86u_p^2)^2}u_p^{3/2}\hat a\nonumber\\
&&
+\frac{u_p^2}{(-1+3u_p)^3(4-39u_p+86u_p^2)^3}(-32+1584u_p-29674u_p^2+297266u_p^3-1801019u_p^4\nonumber\\
&&
+6933366u_p^5-17148037u_p^6+26630422u_p^7-23983868u_p^8+9746632u_p^9){\hat a}^2\nonumber\\
&&
+\bigg\{
-\frac12\frac{(-8+256u_p-3533u_p^2+26468u_p^3-112802u_p^4+265855u_p^5-310004u_p^6+128604u_p^7)}{(-1+3u_p)^2(4-39u_p+86u_p^2)^2(-1+2u_p)(-1+6u_p)}u_p^2\nonumber\\
&&
+\frac14\frac{u_p^{5/2}}{(-1+3u_p)^3(4-39u_p+86u_p^2)^3(-1+2u_p)(-1+6u_p)^2}(-480+22136u_p-463312u_p^2\nonumber\\
&&
+5780997u_p^3-47514985u_p^4+268657357u_p^5-1061333235u_p^6+2919149266u_p^7\nonumber\\
&&
-5467300616u_p^8+6641478880u_p^9-4724705136u_p^{10}+1503019296u_p^{11})\hat a\nonumber\\
&&
-\frac14\frac{u_p^3}{(-1+3u_p)^4(4-39u_p+86u_p^2)^4(-1+2u_p)^2(-1+6u_p)^3}(4224-272768u_p+8278968u_p^2\nonumber\\
&&
-156420964u_p^3+2054426153u_p^4-19850358252u_p^5+145756161682u_p^6-829041253480u_p^7\nonumber\\
&&
+3691145926841u_p^8-12912460620256u_p^9+35399969986324u_p^{10}-75321362601016u_p^{11}\nonumber\\
&&
+122013289648128u_p^{12}-145497047785152u_p^{13}+120396791624256u_p^{14}\nonumber\\
&&
-61694700700032u_p^{15}+14720859668736u_p^{16}){\hat a}^2
\bigg\}e^2
+O({\hat a}^3,e^4)
\,,
\end{eqnarray}
respectively.

To regularize the quantity $\Delta\psi$, it is enough to subtract the large-$l$ limit of its PN expansion, i.e.,
\beq
\Delta\psi=\sum_{\ell=0}^{\infty}\left[\frac{1}{2}\left(\Delta\psi^{l,+}+\Delta\psi^{l,-}\right)-B\right]\,,
\eeq
where the left and right contributions are such that $\Delta\psi^{l,+}=\Delta\psi^{-l-1,-}$ and 
\beq
B(u_p,e,\hat a)=B_0(u_p,\hat a)+e^2B_2(u_p,\hat a)+O(e^4)\,,
\eeq
with 
\begin{eqnarray}
B_0(u_p,\hat a)&=&\frac{21}{16}u_p-\frac{201}{128}u_p^2+\frac{529}{1024}u_p^3+\frac{152197}{16384}u_p^4+\frac{17145445}{262144}u_p^5+\frac{886692225}{2097152}u_p^6+\frac{45206277105}{16777216}u_p^7\nonumber\\
&&
+\frac{9204713714385}{536870912}u_p^8+\frac{1875482334818445}{17179869184}u_p^9\nonumber\\
&&
+\left(-\frac{11}{16}u_p^{3/2}+\frac{19}{128}u_p^{5/2}-\frac{3187}{1024}u_p^{7/2}-\frac{897011}{16384}u_p^{9/2}-\frac{119529091}{262144}u_p^{11/2}-\frac{7322895475}{2097152}u_p^{13/2}\right.\nonumber\\
&&\left.
-\frac{434363072475}{16777216}u_p^{15/2}-\frac{101048547627615}{536870912}u_p^{17/2}-\frac{23167337673070755}{17179869184}u_p^{19/2}\right)\hat a
\nonumber\\
&&
+\left(\frac{1}{4}u_p^2-\frac{117}{128}u_p^3+\frac{707}{64}u_p^4+\frac{2138193}{16384}u_p^5+\frac{91049009}{65536}u_p^6+\frac{27330703781}{2097152}u_p^7\right.\nonumber\\
&&\left.
+\frac{15024987805}{131072}u_p^8+\frac{517781575013205}{536870912}u_p^9\right){\hat a}^2
+O({\hat a}^3,u_p^{10})\,,
\end{eqnarray}
and
\begin{eqnarray}
B_2(u_p,\hat a)&=&-\frac{435}{512}u_p^2-\frac{1155}{1024}u_p^3-\frac{352849}{65536}u_p^4-\frac{5100243}{131072}u_p^5-\frac{2456459237}{8388608}u_p^6-\frac{36003649389}{16777216}u_p^7\nonumber\\
&&
-\frac{32713771158557}{2147483648}u_p^8-\frac{451723973383879}{4294967296}u_p^9\nonumber\\
&&
+\left(\frac{605}{512}u_p^{5/2}+\frac{957}{256}u_p^{7/2}+\frac{558367}{65536}u_p^{9/2}+\frac{11521973}{65536}u_p^{11/2}+\frac{17120777051}{8388608}u_p^{13/2}\right.\nonumber\\
&&\left.
+\frac{80347197891}{4194304}u_p^{15/2}+\frac{347929041937283}{2147483648}u_p^{17/2}+\frac{2760514246569789}{2147483648}u_p^{19/2}\right)\hat a\nonumber\\
&&
+\left(-\frac{523}{512}u_p^3+\frac{843}{256}u_p^4+\frac{892255}{65536}u_p^5-\frac{5413651}{16384}u_p^6-\frac{55709086485}{8388608}u_p^7\right.\nonumber\\
&&\left.
-\frac{337847858229}{4194304}u_p^8-\frac{1751794928899397}{2147483648}u_p^9\right){\hat a}^2
+O({\hat a}^3,u_p^{10})\,.
\end{eqnarray}

\end{widetext}

\subsection{Results}

Our final result for the spin precession invariant $\Delta \psi(u_p,e,\hat a)$ reads
\begin{eqnarray}
\Delta \psi(u_p, e, \hat a) &=&\sum_{i,j=0}^\infty e^i {\hat a}^j \Delta \psi^{(e^i,a^j)}(u_p)\nonumber\\
&=& \Delta \psi^{(e^0,a^0)}+e^2 \Delta \psi^{(e^2,a^0)}
\nonumber\\
&+& \hat a \Delta \psi^{(e^0,a^1)}+\hat a^2\Delta \psi^{(e^0,a^2)}+\ldots\nonumber\\
&+& e^2 \hat a\Delta \psi^{(e^2,a^1)}+e^2 \hat a^2\Delta \psi^{(e^2,a^2)}+\ldots\,.
\end{eqnarray}
The spin-independent part has been computed in Refs. \cite{Kavanagh:2017wot,Bini:2018aps} up to the 9PN level, which we raise here to 9.5PN.
The new terms are 
\begin{eqnarray}
\Delta \psi^{(e^0,a^0)}&=&\Delta \psi^{(e^0,a^0)}\vert_{{\rm Ref.} {\mbox {\scriptsize \cite{Bini:2018aps}}}}
+\Delta \psi^{(e^0,a^0)}_{{\rm 9.5 PN}}\,,\nonumber\\
\Delta \psi^{(e^2,a^0)}&=&\Delta \psi^{(e^2,a^0)}\vert_{{\rm Ref.} {\mbox {\scriptsize \cite{Bini:2018aps}}}}
+\Delta \psi^{(e^2,a^0)}_{{\rm 9.5 PN}}\,,
\end{eqnarray}
with
\begin{widetext}
\begin{eqnarray}
\Delta \psi^{(e^0,a^0)}_{{\rm 9.5 PN}}&=&
\left(-\frac{3130119243444996194647}{11453592870720000} -\frac{180728953}{11025}\pi^2+\frac{23055449891}{385875} \gamma+\frac{316521883}{15435} \ln(2)\right.\nonumber\\
&& \left. 
+\frac{6854694417}{85750} \ln(3)+\frac{23055449891}{771750} \ln(u_p)\right) \pi u_p^{19/2}
\,,\nonumber\\
\Delta \psi^{(e^2,a^0)}_{{\rm 9.5 PN}}&=&
\left(-\frac{352741457149881016281557}{61085828643840000} -\frac{53999919103}{176400}\pi^2+\frac{19991310125293}{18522000} \gamma+\frac{264647617121}{35280} \ln(2)\right. \nonumber\\
&& \left.
-\frac{1983214856673}{1372000} \ln(3)-\frac{206298828125}{296352} \ln(5)+\frac{19991310125293}{ 37044000} \ln(u_p)\right) \pi u_p^{19/2}
\,.
\end{eqnarray}

The zero-eccentricity spin-dependent terms are given by
\begin{eqnarray}
\Delta \psi^{(e^0,a^1)}&=&C_{1.5}^{(e^0,a^1),{\rm c}}  u_p^{3/2}+C_{2.5}^{(e^0,a^1),{\rm c}}  u_p^{5/2}+C_{3.5}^{(e^0,a^1),{\rm c}}  u_p^{7/2}
+\left(C_{4.5}^{(e^0,a^1),{\rm c}}+ C_{4.5}^{{(e^0,a^1)},\ln{}}\ln(u_p) \right) u_p^{9/2}\nonumber\\
&& 
+\left(C_{5.5}^{(e^0,a^1),{\rm c}}+ C_{5.5}^{{(e^0,a^1)},\ln{}} \ln(u_p)\right) u_p^{11/2}+C_{6}^{(e^0,a^1),{\rm c}}  u_p^6
+\left(C_{6.5}^{(e^0,a^1),{\rm c}}+ C_{6.5}^{{(e^0,a^1)}\,\ln{}} \ln(u_p) \right) u_p^{13/2}\nonumber\\
&& +C_{7}^{(e^0,a^1),{\rm c}}  u_p^7
+\left( C_{7.5}^{(e^0,a^1),{\rm c}}+ C_{7.5}^{{(e^0,a^1)},\ln{}}\ln(u_p)+ C_{7.5}^{{(e^0,a^1)},\ln^2{}}\ln(u_p)^2 \right) u_p^{15/2}
+C_{8}^{(e^0,a^1),{\rm c}}  u_p^8\nonumber\\
&&+\left(C_{8.5}^{(e^0,a^1),{\rm c}}+ C_{8.5}^{{(e^0,a^1)},\ln{}} \ln(u_p)+ C_{8.5}^{{(e^0,a^1)},\ln^2{}}\ln(u_p)^2 \right) u_p^{17/2}
+\left(C_{9}^{(e^0,a^1),{\rm c}}+ C_{9}^{{(e^0,a^1)},\ln{}}\ln(u_p) \right) u_p^9\nonumber\\
&&+\left(C_{9.5}^{(e^0,a^1),{\rm c}}+ C_{9.5}^{{(e^0,a^1)},\ln{}} \ln(u_p)+ C_{9.5}^{{(e^0,a^1)},\ln^2{}}\ln(u_p)^2 \right) u_p^{19/2}
+O_{\rm ln}(u_p^{10})
\,,
\end{eqnarray}
with
\begin{eqnarray}
C_{1.5}^{(e^0,a^1),{\rm c}}&=&-\frac12  
\,,\qquad 
C_{2.5}^{(e^0,a^1),{\rm c}}=-\frac{41}{8}
\,,\qquad  
C_{3.5}^{(e^0,a^1),{\rm c}}=\frac{237}{32}-\frac{123}{64}\pi^2
\,,\nonumber\\
C_{4.5}^{(e^0,a^1),{\rm c}}&=&-\frac{2580077}{5760}+\frac{52225}{6144}\pi^2+\frac{1256}{15}\gamma +\frac{296}{15}\ln(2)+\frac{729}{5}\ln(3)
\,,\qquad
C_{4.5}^{{(e^0,a^1)},\ln{}}= \frac{628}{15}
\,,\nonumber\\
C_{5.5}^{(e^0,a^1),{\rm c}}&=&-\frac{371061}{140}\ln(3)+\frac{16521221}{24576}\pi^2-\frac{653849867}{115200}+\frac{20186}{35}\ln(2)-\frac{131234}{105}\gamma 
\,,\nonumber\\
C_{5.5}^{{(e^0,a^1)},\ln{}} &=&-\frac{65617}{105}  
\,,\qquad
C_{6}^{(e^0,a^1),{\rm c}} =\frac{49969}{315}\pi 
\,,\nonumber\\
C_{6.5}^{(e^0,a^1),{\rm c}}&=& -\frac{34667196284353}{203212800}+\frac{43396897187}{2359296}\pi^2+\frac{4274383}{1890}\gamma-\frac{5127317}{378}\ln(2)+\frac{602397}{70}\ln(3)\nonumber\\
&& 
+\frac{9765625}{9072}\ln(5)-\frac{7335303}{131072}\pi^4
\,,\qquad
C_{6.5}^{{(e^0,a^1)},\ln{}}= \frac{4111087}{3780}  
\,,\qquad
C_{7}^{(e^0,a^1),{\rm c}}   = -\frac{17884343}{6300}\pi
\,,\nonumber\\
C_{7.5}^{(e^0,a^1),{\rm c}} &=& -\frac{530755103526042557}{521579520000}+\frac{138120741638137}{1238630400}\pi^2+\frac{1796383502593}{43659000}\gamma+\frac{7478658446233}{43659000}\ln(2)\nonumber\\
&&
+\frac{60948732447}{8624000}\ln(3) -\frac{2216796875}{72576}\ln(5)-\frac{1951932086423}{1006632960}\pi^4+\frac{63488}{15}\zeta(3)-\frac{3396608}{1575}\gamma^2\nonumber\\
&&
 -\frac{5149696}{1575}\gamma\ln(2)-\frac{936036}{175}\gamma\ln(3)-\frac{931328}{1575}\ln(2)^2-\frac{936036}{175}\ln(3)\ln(2) -\frac{468018}{175}\ln(3)^2
\,,\nonumber\\
C_{7.5}^{{(e^0,a^1)},\ln{}}&=& \frac{1781539442593}{87318000}  -\frac{3396608}{1575}\gamma -\frac{2574848}{1575}\ln(2) -\frac{468018}{175} \ln(3)
\,,\qquad
C_{7.5}^{{(e^0,a^1)},\ln^2{}}= -\frac{849152}{1575}
\,,\nonumber\\
C_{8}^{(e^0,a^1),{\rm c}} &=&\frac{27936275503}{2910600}\pi 
\,,\nonumber\\
C_{8.5}^{(e^0,a^1),{\rm c}} &=& -\frac{681266651719214562649}{11277383749632000}+\frac{427384464568822843}{1109812838400}\pi^2-\frac{5947119623686361}{15891876000}\gamma
\nonumber\\
&&
-\frac{5484803096524561}{15891876000}\ln(2)-\frac{2417456672510751}{3139136000}\ln(3)+\frac{72943791015625}{290594304}\ln(5)\nonumber\\
&&
+\frac{678223072849}{92664000}\ln(7)-\frac{466989768838667}{12884901888}\pi^4-\frac{6177152}{105}\zeta(3)+\frac{46649968}{1225}\gamma^2+\frac{30413792}{1225}\gamma\ln(2)\nonumber\\
&&
+\frac{137032317}{1225}\gamma\ln(3)-\frac{599506832}{11025}\ln(2)^2+\frac{137032317}{1225}\ln(3)\ln(2) +\frac{137032317}{2450}\ln(3)^2
\,,\nonumber\\
C_{8.5}^{{(e^0,a^1)},\ln{}}&=& -\frac{5994748484165561}{31783752000}+\frac{46649968}{1225}\gamma  +\frac{15206896}{1225}\ln(2) +\frac{137032317}{2450} \ln(3)
\,,\qquad
C_{8.5}^{{(e^0,a^1)},\ln^2{}}= \frac{11662492}{1225} 
\,,\nonumber
\end{eqnarray}
\begin{eqnarray}
C_{9}^{(e^0,a^1),{\rm c}} &=&\frac{116182866505170823}{2097727632000}\pi+\frac{10999172}{4725}\pi^3-\frac{1176911404}{165375}\pi\gamma-\frac{1001054764}{165375}\pi\ln(2)   -\frac{50077926}{6125}\pi\ln(3)
\,,\nonumber\\
C_{9}^{{(e^0,a^1)},\ln{}}&=&-\frac{588455702}{165375}\pi 
\,,\nonumber\\
C_{9.5}^{(e^0,a^1),{\rm c}}&=& \frac{209166701047899777145941874951}{290279857715527680000}+\frac{458671024959899491}{775107379200}\pi^2+ \frac{9661311739976843}{419545526400}\gamma\nonumber\\
&& 
-\frac{50685948367874925289}{18879548688000}\ln(2)+\frac{69137888408251731}{27624396800}\ln(3)-\frac{4945799701746484375}{7249746696192}\ln(5)\nonumber\\
&&
-\frac{102411684000199}{370656000}\ln(7)-\frac{49177882351749818983}{6597069766656}\pi^4+\frac{1719608}{63}\zeta(3) -\frac{131422044664}{1091475}\gamma^2\nonumber\\
&&
+\frac{1715126780656}{3274425}\gamma\ln(2)-\frac{7539716034}{13475}\gamma\ln(3)-\frac{76708984375}{1571724}\gamma\ln(5)+\frac{14049792489496}{9823275}\ln(2)^2\nonumber\\
&& 
-\frac{7539716034}{13475}\ln(3)\ln(2)-\frac{76708984375}{1571724}\ln(2)\ln(5)-\frac{3769858017}{13475}\ln(3)^2 \nonumber\\
&&
-\frac{76708984375}{3143448}\ln(5)^2+\frac{128148402261}{67108864}\pi^6 
\,,\nonumber\\
C_{9.5}^{{(e^0,a^1)},\ln{}}&=& -\frac{4238891698612841}{4195455264000} -\frac{131422044664}{1091475}\gamma+\frac{863777837336}{3274425}\ln(2)-\frac{3769858017}{13475} \ln(3) -\frac{76708984375}{3143448}\ln(5)
\,,\nonumber\\
C_{9.5}^{{(e^0,a^1)},\ln^2{}}&=& -\frac{31819769998}{1091475} 
\,,
\end{eqnarray}
and
\begin{eqnarray}
\Delta \psi^{(e^0,a^2)}&=&C_{2}^{(e^0,a^2),{\rm c}}  u_p^{2}+C_{3}^{(e^0,a^2),{\rm c}}  u_p^{3}+C_{4}^{(e^0,a^2),{\rm c}}  u_p^{4}+\left(C_{5}^{(e^0,a^2),{\rm c}}+ C_{5}^{{(e^0,a^2)},\ln{}}\ln(u_p) \right) u_p^{5}\nonumber\\
&&
+\left(C_{6}^{(e^0,a^2),{\rm c}}+ C_{6}^{{(e^0,a^2)},\ln{}} \ln(u_p)\right) u_p^{6}
 +C_{6.5}^{(e^0,a^2),{\rm c}}u_p^{13/2}\nonumber\\
&&
+\left( C_{7}^{(e^0,a^2),{\rm c}}+ C_{7}^{{(e^0,a^2)},\ln{}}\ln(u_p)\right) u_p^{7}
+C_{7.5}^{(e^0,a^2),{\rm c}}  u_p^{15/2}\nonumber\\
&&
+\left(C_{8}^{(e^0,a^2),{\rm c}}+ C_{8}^{{(e^0,a^2)},\ln{}} \ln(u_p)+ C_{8}^{{(e^0,a^2)},\ln^2{}}\ln(u_p)^2 \right) u_p^8
+C_{8.5}^{(e^0,a^2),{\rm c}} u_p^{17/2}\nonumber\\
&&
+\left(C_{9}^{(e^0,a^2),{\rm c}}+ C_{9}^{{(e^0,a^2)},\ln{}} \ln(u_p)+ C_{9}^{{(e^0,a^2)},\ln^2{}}\ln(u_p)^2 \right)u_p^9\nonumber\\
&&
+\left(C_{9.5}^{(e^0,a^2),{\rm c}}+ C_{9.5}^{{(e^0,a^2)},\ln{}} \ln(u_p)\right) u_p^{19/2}
+O_{\rm ln}(u_p^{10})
\,,
\end{eqnarray}
with
\begin{eqnarray}
C_{2}^{(e^0,a^2),{\rm c}}&=&-1\,,\qquad  
C_{3}^{(e^0,a^2),{\rm c}}= \frac{15}{4}\,,\qquad  
C_{4}^{(e^0,a^2),{\rm c}}=\frac{843}{16}-\frac{123}{64}\pi^2 
\,,\nonumber\\
C_{5}^{(e^0,a^2),{\rm c}}&=&-\frac{41161}{2880}+\frac{5155}{1536}\pi^2+\frac{1256}{15}\gamma+\frac{296}{15}\ln(2)+\frac{729}{5}\ln(3) 
\,,\qquad
C_{5}^{{(e^0,a^2)},\ln{}}=  \frac{628}{15} 
\,,\nonumber\\
C_{6}^{(e^0,a^2),{\rm c}}&=&-\frac{198163141}{57600}+\frac{14769449}{24576}\pi^2-\frac{32484}{35}\gamma+\frac{53012}{105}\ln(2)-\frac{68526}{35}\ln(3) 
\,,\qquad
C_{6}^{{(e^0,a^2)},\ln{}}=-\frac{16242}{35}  
\,,\nonumber\\
C_{6.5}^{(e^0,a^2),{\rm c}}&=& \frac{49969}{315}\pi
\,,\nonumber\\
C_{7}^{(e^0,a^2),{\rm c}}&=&-\frac{31986710669261}{101606400}+\frac{79045202729}{2359296}\pi^2+\frac{8403943}{1890}\gamma-\frac{35370913}{1890}\ln(2) +\frac{5276259}{280}\ln(3)\nonumber\\
&&
+\frac{9765625}{9072}\ln(5) -\frac{7335303}{131072}\pi^4
\,,\qquad
C_{7}^{{(e^0,a^2)},\ln{}}= \frac{8240647}{3780}  
\,,\nonumber\\
C_{7.5}^{(e^0,a^2),{\rm c}}&=&-\frac{113991}{50}\pi 
\,,\nonumber\\
C_{8}^{(e^0,a^2),{\rm c}}&=&-\frac{1336810434105217691}{260789760000}+\frac{876841593090859}{1651507200}\pi^2+\frac{24691487069}{606375}\gamma +\frac{241419814667}{1819125}\ln(2)\nonumber\\
&&  
+\frac{367133665347}{8624000}\ln(3)-\frac{1689453125}{72576}\ln(5)-\frac{222475429201}{125829120}\pi^4+\frac{89552}{15}\zeta(3)-\frac{3396608}{1575}\gamma^2\nonumber\\
&&
-\frac{5149696}{1575}\gamma\ln(2)-\frac{936036}{175}\gamma\ln(3)-\frac{931328}{1575}\ln(2)^2-\frac{936036}{175}\ln(3)\ln(2)-\frac{468018}{175}\ln(3)^2
\,,\nonumber
\end{eqnarray}
\begin{eqnarray}
C_{8}^{{(e^0,a^2)},\ln{}}&=& \frac{26855194769}{1212750}-\frac{3396608}{1575}\gamma -\frac{2574848}{1575}\ln(2) -\frac{468018}{175} \ln(3) 
\,,\qquad
C_{8}^{{(e^0,a^2)},\ln^2{}}= -\frac{849152}{1575} 
\,,\nonumber\\
C_{8.5}^{(e^0,a^2),{\rm c}}&=&\frac{2454668003}{138600}\pi 
\,,\nonumber\\
C_{9}^{(e^0,a^2),{\rm c}}&=& -\frac{576916351095208258869353}{28193459374080000}+\frac{1743178610265080953}{554906419200}\pi^2-\frac{4986393924529319}{15891876000}\gamma\nonumber\\
&&
-\frac{766147036782487}{635675040}\ln(2)-\frac{208971977923053}{1569568000}\ln(3)  +\frac{23691548828125}{72648576}\ln(5)+\frac{678223072849}{92664000}\ln(7)\nonumber\\
&&
-\frac{2174958156794893}{21474836480}\pi^4-\frac{290740}{7}\zeta(3)+\frac{344032936}{11025}\gamma^2+\frac{192844112}{11025}\gamma\ln(2)+\frac{112461372}{1225}\gamma\ln(3)\nonumber\\
&&
-\frac{569193304}{11025}\ln(2)^2+\frac{112461372}{1225}\ln(3)\ln(2)+\frac{56230686}{1225}\ln(3)^2
\,,\nonumber\\
C_{9}^{{(e^0,a^2)},\ln{}} &=& -\frac{4888595168274119}{31783752000} +\frac{344032936}{11025}\gamma+\frac{96422056}{11025}\ln(2) +\frac{56230686}{1225} \ln(3)
\,, \qquad
C_{9}^{{(e^0,a^2)},\ln^2{}}=   \frac{86008234}{11025} 
\,,\nonumber\\
C_{9.5}^{(e^0,a^2),{\rm c}}&=& \frac{69193216974146623}{2097727632000}\pi+\frac{10999172}{4725}\pi^3-\frac{1176911404}{165375}\pi\gamma-\frac{1001054764}{165375}\pi\ln(2)-\frac{50077926}{6125}\pi\ln(3) 
\,,\nonumber\\
C_{9.5}^{{(e^0,a^2)},\ln{}}&=&-\frac{588455702}{165375}\pi
\,.
\end{eqnarray}

Finally, the spin-dependent part mixing eccentricity and spin effects is given by
\begin{eqnarray}
\Delta \psi^{(e^2,a^1)}&=&C_{2.5}^{(e^2,a^1),{\rm c}}  u_p^{5/2}+C_{3.5}^{(e^2,a^1),{\rm c}}  u_p^{7/2}+\left(C_{4.5}^{(e^2,a^1),{\rm c}}+ C_{4.5}^{{(e^2,a^1)},\ln{}}\ln(u_p) \right) u_p^{9/2}\nonumber\\
&&+\left(C_{5.5}^{(e^2,a^1),{\rm c}}+ C_{5.5}^{{(e^2,a^1)},\ln{}} \ln(u_p)\right) u_p^{11/2}
 +C_{6}^{(e^2,a^1),{\rm c}}  u_p^6
 +\left(C_{6.5}^{(e^2,a^1),{\rm c}}+ C_{6.5}^{{(e^2,a^1)},\ln{}} \ln(u_p) \right) u_p^{13/2}\nonumber\\
&& +C_{7}^{(e^2,a^1),{\rm c}}  u_p^7
+\left( C_{7.5}^{(e^2,a^1),{\rm c}}+ C_{7.5}^{{(e^2,a^1)},\ln{}}\ln(u_p)+ C_{7.5}^{{(e^2,a^1)},\ln^2{}}\ln(u_p)^2 \right) u_p^{15/2}
+C_{8}^{(e^2,a^1),{\rm c}}  u_p^8\nonumber\\
&&+\left(C_{8.5}^{(e^2,a^1),{\rm c}}+ C_{8.5}^{{(e^2,a^1)},\ln{}} \ln(u_p)+ C_{8.5}^{{(e^2,a^1)},\ln^2{}}\ln(u_p)^2 \right) u_p^{17/2}
+\left(C_{9}^{(e^2,a^1),{\rm c}}+ C_{9}^{{(e^2,a^1)},\ln{}}\ln(u_p) \right) u_p^9\nonumber\\
&&
+\left(C_{9.5}^{(e^2,a^1),{\rm c}}+ C_{9.5}^{{(e^2,a^1)},\ln{}} \ln(u_p)+ C_{9.5}^{{(e^2,a^1)},\ln^2{}}\ln(u_p)^2 \right) u_p^{19/2}
+O_{\rm ln}(u_p^{10})
\,,
\end{eqnarray}
with
\begin{eqnarray}
C_{2.5}^{(e^2,a^1),{\rm c}}&=& -\frac18    
\,,\qquad C_{3.5}^{(e^2,a^1),{\rm c}}=-\frac{59}{16}-\frac{123}{256}\pi^2   
\,,\nonumber\\
C_{4.5}^{(e^2,a^1),{\rm c}}&=&-\frac{274889}{640}-\frac{39529}{4096}\pi^2+\frac{536}{5}\gamma+\frac{11720}{3}\ln(2) -\frac{10206}{5}\ln(3) 
\,,\qquad
C_{4.5}^{{(e^2,a^1)},\ln{}}=\frac{268}{5}  
\,,\nonumber\\
C_{5.5}^{(e^2,a^1),{\rm c}}&=& -\frac{47376713}{14400}+\frac{46450919}{49152}\pi^2-\frac{38026}{15}\gamma-\frac{2049574}{21}\ln(2)+\frac{13574709}{320}\ln(3)+\frac{9765625}{1344}\ln(5) 
\,,\nonumber\\
C_{5.5}^{{(e^2,a^1)},\ln{}} &=& -\frac{19013}{15} 
\,,\qquad
C_{6}^{(e^2,a^1),{\rm c}} = \frac{319609}{630}\pi  
\,,\nonumber\\
C_{6.5}^{(e^2,a^1),{\rm c}}&=& -\frac{19506870722893}{29030400}+\frac{168336760679}{2359296}\pi^2+\frac{187867}{27}\gamma+\frac{616924811}{945}\ln(2)-\frac{111860433}{1792}\ln(3)\nonumber\\
&& 
-\frac{4701171875}{20736}\ln(5)-\frac{146026515}{1048576}\pi^4
\,,\qquad
C_{6.5}^{{(e^2,a^1)},\ln{}}=  \frac{179443}{54}   
\,,\nonumber\\
C_{7}^{(e^2,a^1),{\rm c}}   &=&-\frac{629699771}{47040}\pi 
\,,\nonumber\\
C_{7.5}^{(e^2,a^1),{\rm c}} &=& -\frac{6966671370033684457}{782369280000}+\frac{6083404435612271}{6606028800}\pi^2+\frac{4282750559249}{14553000}\gamma-\frac{23584521073621}{8731800}\ln(2)\nonumber\\
&& 
-\frac{271718217011673}{275968000}\ln(3)+\frac{56509677734375}{25546752}\ln(5)+\frac{678223072849}{6082560}\ln(7)-\frac{3390769890109}{335544320}\pi^4\nonumber\\
&& 
+\frac{134944}{5}\zeta(3)-\frac{7219504}{525}\gamma^2-\frac{79652512}{315}\gamma\ln(2)+\frac{15912612}{175}\gamma\ln(3)-\frac{80263696}{175}\ln(2)^2 \nonumber\\
&&
+\frac{15912612}{175}\ln(3)\ln(2)+\frac{7956306}{175}\ln(3)^2
\,,\nonumber\\
C_{7.5}^{{(e^2,a^1)},\ln{}}&=&  \frac{4253184684449}{29106000} -\frac{7219504}{525}\gamma -\frac{39826256}{315}\ln(2) +\frac{7956306}{175} \ln(3) 
\,,\qquad
C_{7.5}^{{(e^2,a^1)},\ln^2{}}= -\frac{1804876}{525} 
\,,\nonumber\\
C_{8}^{(e^2,a^1),{\rm c}} &=& \frac{21791194144427}{279417600}\pi 
\,,\nonumber
\end{eqnarray}
\begin{eqnarray}
C_{8.5}^{(e^2,a^1),{\rm c}} &=&-\frac{335314468446423168647897}{11277383749632000}+\frac{10173198456270880813}{2219625676800}\pi^2 -\frac{27456615049027529}{7945938000}\gamma\nonumber\\
&&
-\frac{170592889938737863}{7945938000}\ln(2)+\frac{530521027102134009}{20090470400}\ln(3)-\frac{558244786663203125}{65093124096}\ln(5)\nonumber\\
&&
-\frac{106301099544967201}{23721984000}\ln(7)-\frac{31433138995123013}{257698037760}\pi^4-\frac{49939936}{105}\zeta(3)+\frac{1202995352}{3675}\gamma^2\nonumber\\
&&
+\frac{15075112912}{2205}\gamma\ln(2)-\frac{41368207977}{19600}\gamma\ln(3)-\frac{3173828125}{7056}\gamma\ln(5)+\frac{138535911176}{11025}\ln(2)^2\nonumber\\
&&
-\frac{41368207977}{19600}\ln(3)\ln(2)-\frac{3173828125}{7056}\ln(2)\ln(5)-\frac{41368207977}{39200}\ln(3)^2-\frac{3173828125}{14112}\ln(5)^2 
\,,\nonumber\\
C_{8.5}^{{(e^2,a^1)},\ln{}}&=& -\frac{27648022365728129}{15891876000}+\frac{1202995352}{3675}\gamma +\frac{7537556456}{2205}\ln(2)  -\frac{41368207977}{39200} \ln(3)  -\frac{3173828125}{14112}\ln(5)
\,,\nonumber\\
C_{8.5}^{{(e^2,a^1)},\ln^2{}}&=&  \frac{300748838}{3675}  
\,,\nonumber\\
C_{9}^{(e^2,a^1),{\rm c}} &=&\frac{21196624539283901299}{33563642112000}\pi+\frac{123567238}{4725}\pi^3-\frac{13221694466}{165375}\pi\gamma -\frac{166628746}{315}\pi\ln(2)+\frac{926441631}{6125}\pi\ln(3)
\,,\nonumber\\
C_{9}^{{(e^2,a^1)},\ln{}}&=&-\frac{6610847233}{165375}\pi  
\,,\nonumber\\
C_{9.5}^{(e^2,a^1),{\rm c}}&=&\frac{39739290623493246364083403421}{6597269493534720000}+\frac{73642835756807424659}{8878502707200}\pi^2+\frac{151919323439718707}{27243216000}\gamma\nonumber\\
&& 
+\frac{82978352154780707393}{699242544000}\ln(2)-\frac{23067310031610488793}{339992576000}\ln(3)-\frac{16664548214107826421875}{231991894278144}\ln(5)\nonumber\\
&&
+\frac{16814208792873724897}{284663808000}\ln(7)-\frac{4121605749668435649521}{65970697666560}\pi^4+\frac{27893212}{45}\zeta(3)-\frac{25587350981}{14175}\gamma^2\nonumber\\
&&
-\frac{63046858231586}{1091475}\gamma\ln(2)+\frac{1269422541711}{172480}\gamma\ln(3)+\frac{374473583984375}{25147584}\gamma\ln(5)\nonumber\\
&&
-\frac{1074807834943193}{9823275}\ln(2)^2+\frac{931671919503}{172480}\ln(3)\ln(2)+\frac{374473583984375}{25147584}\ln(2)\ln(5)\nonumber\\
&&
+\frac{1269422541711}{344960}\ln(3)^2+\frac{374473583984375}{50295168}\ln(5)^2  +\frac{962681186487}{268435456}\pi^6
\,,\nonumber\\
C_{9.5}^{{(e^2,a^1)},\ln{}}&=& \frac{145588636751786243}{54486432000}-\frac{25587350981}{14175}\gamma  -\frac{1259924060281}{43659}\ln(2)+\frac{1269422541711}{344960} \ln(3)\nonumber\\
&& +\frac{374473583984375}{50295168}\ln(5)   
\,,\qquad
C_{9.5}^{{(e^2,a^1)},\ln^2{}}= -\frac{3561355859}{8100} 
\,,
\end{eqnarray}
and
\begin{eqnarray}
\Delta \psi^{(e^2,a^2)}&=&C_{3}^{(e^2,a^2),{\rm c}}  u_p^{3}+C_{4}^{(e^2,a^2),{\rm c}}  u_p^{4}+\left(C_{5}^{(e^2,a^2),{\rm c}}+ C_{5}^{{(e^2,a^2)},\ln{}}\ln(u_p) \right) u_p^{5}\nonumber\\
&&
+\left(C_{6}^{(e^2,a^2),{\rm c}}+ C_{6}^{{(e^2,a^2)},\ln{}} \ln(u_p)\right) u_p^{6}
 +C_{6.5}^{(e^2,a^2),{\rm c}}u_p^{13/2}\nonumber\\
&&
+\left( C_{7}^{(e^2,a^2),{\rm c}}+ C_{7}^{{(e^2,a^2)},\ln{}}\ln(u_p)\right) u_p^{7}
+C_{7.5}^{(e^2,a^2),{\rm c}}  u_p^{15/2}\nonumber\\
&&
+\left(C_{8}^{(e^2,a^2),{\rm c}}+ C_{8}^{{(e^2,a^2)},\ln{}} \ln(u_p)+ C_{8}^{{(e^2,a^2)},\ln^2{}}\ln(u_p)^2 \right) u_p^8
+C_{8.5}^{(e^2,a^2),{\rm c}} u_p^{17/2}\nonumber\\
&&
+\left(C_{9}^{(e^2,a^2),{\rm c}}+ C_{9}^{{(e^2,a^2)},\ln{}} \ln(u_p)+ C_{9}^{{(e^2,a^2)},\ln^2{}}\ln(u_p)^2 \right)u_p^9\nonumber\\
&&
+\left(C_{9.5}^{(e^2,a^2),{\rm c}}+ C_{9.5}^{{(e^2,a^2)},\ln{}} \ln(u_p)\right) u_p^{19/2}
+O_{\rm ln}(u_p^{10})
\,,
\end{eqnarray}
with
\begin{eqnarray}
C_{3}^{(e^2,a^2),{\rm c}}&=&-2 
\,,\qquad  
C_{4}^{(e^2,a^2),{\rm c}}= -\frac{13}{4}-\frac{123}{256}\pi^2
\,,\nonumber\\ 
C_{5}^{(e^2,a^2),{\rm c}}&=&-\frac{65091}{160}-\frac{22037}{2048}\pi^2+\frac{536}{5}\gamma+\frac{11720}{3}\ln(2) -\frac{10206}{5}\ln(3) 
\,,\qquad
C_{5}^{{(e^2,a^2)},\ln{}}=  \frac{268}{5} 
\,,\nonumber\\
C_{6}^{(e^2,a^2),{\rm c}}&=&-\frac{9371747}{3600}+\frac{33970805}{49152}\pi^2-\frac{31018}{15}\gamma-\frac{8107718}{105}\ln(2)+\frac{10023021}{320}\ln(3)+\frac{9765625}{1344}\ln(5)
\,,\nonumber\\
C_{6}^{{(e^2,a^2)},\ln{}}&=& -\frac{15509}{15}   
\,,\qquad
C_{6.5}^{(e^2,a^2),{\rm c}}= \frac{319609}{630}\pi
\,,\nonumber
\end{eqnarray}
\begin{eqnarray}
C_{7}^{(e^2,a^2),{\rm c}}&=& -\frac{9278011192573}{7257600}+\frac{308714314565}{2359296}\pi^2+\frac{18377071}{1890}\gamma+\frac{1961637691}{1890}\ln(2)-\frac{39782259}{112}\ln(3) \nonumber\\
&&
-\frac{3302734375}{18144}\ln(5)-\frac{146026515}{1048576}\pi^4
\,,\qquad
C_{7}^{{(e^2,a^2)},\ln{}}= \frac{17787391}{3780}    
\,,\qquad
C_{7.5}^{(e^2,a^2),{\rm c}}=-\frac{916708909}{78400}\pi  
\,,\nonumber\\
C_{8}^{(e^2,a^2),{\rm c}}&=&-\frac{7515099422720578033}{195592320000}+\frac{2867592560250501}{734003200}\pi^2+\frac{3469177106249}{14553000}\gamma-\frac{19693190989441}{8731800}\ln(2) \nonumber\\
&& 
-\frac{625767867535473}{275968000}\ln(3) +\frac{3414326171875}{1216512}\ln(5)+\frac{678223072849}{6082560}\ln(7)-\frac{365600352653}{41943040}\pi^4\nonumber\\
&&
+\frac{196096}{5}\zeta(3)-\frac{7219504}{525}\gamma^2-\frac{79652512}{315}\gamma\ln(2) +\frac{15912612}{175}\gamma\ln(3)-\frac{80263696}{175}\ln(2)^2\nonumber\\
&& 
+\frac{15912612}{175}\ln(2)\ln(3) +\frac{7956306}{175}\ln(3)^2
\,,\nonumber\\
C_{8}^{{(e^2,a^2)},\ln{}}&=& \frac{3819871419449}{29106000} -\frac{7219504}{525}\gamma -\frac{39826256}{315}\ln(2) + \frac{7956306}{175} \ln(3) 
\,,\qquad
C_{8}^{{(e^2,a^2)},\ln^2{}}=  -\frac{1804876}{525}  
\,,\nonumber\\
C_{8.5}^{(e^2,a^2),{\rm c}}&=& \frac{16093843572391}{139708800}\pi 
\,,\nonumber\\
C_{9}^{(e^2,a^2),{\rm c}}&=& -\frac{937636103996199692927831}{2819345937408000}+\frac{5639207064852757987}{138726604800}\pi^2-\frac{57750252464296303}{15891876000}\gamma \nonumber\\
&&
+\frac{217123592856512149}{15891876000}\ln(2)+\frac{127161994962779433}{12556544000}\ln(3)-\frac{116148004073984375}{8136640512}\ln(5)  \nonumber\\
&&
-\frac{478922378441801}{134784000}\ln(7) -\frac{167245560394225319}{257698037760}\pi^4-\frac{39051856}{105}\zeta(3)+\frac{358048294}{1225}\gamma^2\nonumber\\
&&
+\frac{13108799164}{2205}\gamma\ln(2)-\frac{1359614889}{784}\gamma\ln(3)-\frac{3173828125}{7056}\gamma\ln(5) +\frac{24087679582}{2205}\ln(2)^2\nonumber\\
&&
-\frac{1359614889}{784}\ln(2)\ln(3)-\frac{3173828125}{7056}\ln(2)\ln(5)-\frac{1359614889}{1568}\ln(3)^2-\frac{3173828125}{14112}\ln(5)^2 
\,,\nonumber\\
C_{9}^{{(e^2,a^2)},\ln{}} &=& -\frac{57076219203195103}{31783752000}+\frac{358048294}{1225}\gamma +\frac{6554399582}{2205}\ln(2) -\frac{1359614889}{1568} \ln(3) -\frac{3173828125}{14112} \ln(5) 
\,,\nonumber\\
C_{9}^{{(e^2,a^2)},\ln^2{}}&=& \frac{179024147}{2450}  
\,,\nonumber\\
C_{9.5}^{(e^2,a^2),{\rm c}}&=&  \frac{8587514628160355479}{33563642112000}\pi +\frac{123567238}{4725}\pi^3-\frac{13221694466}{165375}\pi\gamma-\frac{166628746}{315}\pi\ln(2)+\frac{926441631}{6125}\pi\ln(3)
\,,\nonumber\\
C_{9.5}^{{(e^2,a^2)},\ln{}}&=&-\frac{6610847233}{165375}\pi 
\,.
\end{eqnarray}
\end{widetext}

The structure of the first PN terms shows an interesting resummation property, which has been discussed in Ref. \cite{Bini2019:deltakcirc} (see Eq. (6) there).

\subsection{Circular orbit limit}

Let us consider now the zero-eccentricity limit of the above expressions.
In the non-spinning case Akcay et al. \cite{Akcay:2016dku} showed that the difference between the limit for vanishing eccentricity of $\Delta\psi$, i.e., lim$_{e\to0}\Delta\psi$, and the corresponding quantity $\Delta\psi^{\rm circ}$ calculated for circular orbits is proportional to the SF correction to the fractional periastron advance, which is fully known up to the 9.5PN order in terms of the EOB function $\rho$ \cite{Damour:2009sm,Bini:2016qtx}.
The same functional relation has been argued to hold in the Kerr case \cite{Akcay:2017azq}, even if the gauge-invariant SF correction to the periastron advance for
circular equatorial orbits in a Kerr spacetime is not explicitly known, i.e.,
\beq
\label{circlimdeltapsi}
\lim_{e\to0}\Delta\psi-\Delta\psi^{\rm circ}=\bar G_\psi \Delta k\,,
\eeq
where
\beq
2\pi\Delta k=\Delta\Phi\vert_{e\to0}
=\delta\Phi\vert_{e\to0}-\frac{\partial \bar \Phi^{\rm circ}}{\partial \bar\Omega_\phi^{\rm circ}}\delta\Omega_\phi\vert_{e\to0}\,,
\eeq
and
\beq
\bar G_\psi=-\frac{2\pi}{\bar g_1}\frac{\partial \bar \psi}{\partial \bar\Omega_r}\,,\qquad
\bar g_1=-\frac1{2\pi}\bar T_r\bar\Phi\vert_{e\to0}\,,
\eeq
which turns out to be

\begin{widetext}
\begin{eqnarray}
\bar G_\psi&=& -\frac{2(1-6u_p)^{5/2}(1-3u_p)^{1/2}}{(86u_p^2-39u_p+4)}
+\frac{2u_p^{1/2}  (1-6u_p)^{3/2}}{ (86u_p^2-39u_p+4)^2 (1-3u_p)^{1/2}}(744u_p^4-384u_p^3-28u_p^2+37u_p-4)\, \hat a\nonumber\\
&-& \frac{u_p (1-6u_p)^{1/2}}{ (86u_p^2-39u_p+4)^3 (1-3u_p)^{3/2}}
(794376u_p^8-2135148u_p^7+2333418u_p^6-1376961u_p^5\nonumber\\
&& 
+484745u_p^4-105147u_p^3+13836u_p^2-1016u_p+32)\, {\hat a}^2
+O({\hat a}^3)
\,,
\end{eqnarray}
\end{widetext}
to the second order in the rotation parameter.
Therefore, one needs to compute also the GSF corrections \eqref{deltaTePhi} to the periods and \eqref{deltaOmegas} to the associated frequencies. 

The correction $\Delta\psi^{\rm circ}$ to the spin-precession invariant for circular orbits has been calculated in Ref. \cite{Bini:2018ylh} through the 8PN order and to all orders in spin (see Eq. (4.1)--(4.2) there).
We have checked that Eq. \eqref{circlimdeltapsi} reproduces such result up to the second order in spin.
As a byproduct, we can improve it to the 9.5PN order with the addition of the following new terms
\begin{widetext}
\begin{eqnarray}
\Delta\psi^{{\rm circ}\,, a^1}&=&\Delta\psi^{{\rm circ}\,, a^1}\vert_{{\rm Ref.} {\mbox {\scriptsize \cite{Bini:2018ylh}}}}\nonumber\\
&&
+\left(-\frac{163659814070959}{382016250}+\frac{13576618358917}{309657600}\pi^2+\frac{23552516744}{5457375}\gamma+\frac{1137772376}{218295}\ln(2)+\frac{3826683}{3520}\ln(3)\right. \nonumber\\
&&
+\frac{9765625}{19008}\ln(5)-\frac{3418003793}{67108864}\pi^4+\frac{4064}{5}\zeta(3)-\frac{217424}{525}\gamma^2-\frac{58208}{35}\gamma\ln(2)-\frac{124976}{75}\ln(2)^2\nonumber\\
&&\left.
+\frac{11828649172}{5457375}\ln(y)-\frac{217424}{525}\gamma\ln(y)-\frac{29104}{35}\ln(2)\ln(y)-\frac{54356}{525}\ln(y)^2\right) y^{17/2}\nonumber\\
&&
-\frac{2201017711}{6548850}\pi y^9\nonumber\\
&& 
+\left(\frac{16542752726965594}{1251485235}+\frac{109676435084511079}{1664719257600}\pi^2+\frac{2310004910264}{1489863375}\gamma-\frac{65918048552}{30405375}\ln(2)\right.\nonumber\\
&&
+\frac{6150898410939}{392392000}\ln(3)-\frac{10900390625}{2223936}\ln(5)-\frac{1835842082140957}{12884901888}\pi^4+\frac{30656}{105}\zeta(3)+\frac{492928}{1225}\gamma^2\nonumber\\
&&
+\frac{18697984}{4725}\gamma\ln(2)-\frac{113724}{49}\gamma\ln(3)+\frac{11314048}{2205}\ln(2)^2-\frac{113724}{49}\ln(3)\ln(2)-\frac{56862}{49}\ln(3)^2\nonumber\\
&&
+\frac{1183607831932}{1489863375}\ln(y)+\frac{492928}{1225}\gamma\ln(y)
+\frac{9348992}{4725}\ln(2)\ln(y)-\frac{56862}{49}\ln(y)\ln(3)\nonumber\\
&&\left.
+\frac{123232}{1225}\ln(y)^2\right) y^{19/2}
+O_{\rm ln}(y^{10})\,,
\end{eqnarray}
(linear in the dimensionless spin parameter $\hat a$) and
\begin{eqnarray}
\Delta\psi^{{\rm circ}\,, a^2}&=&\Delta\psi^{{\rm circ}\,, a^2}\vert_{{\rm Ref.} {\mbox {\scriptsize \cite{Bini:2018ylh}}}}\nonumber\\
&&
-\frac{188848}{1575}\pi y^{17/2}
\nonumber\\
&&
+\left(-\frac{3255185322968}{893025}+\frac{231004545858251}{619315200}\pi^2+\frac{1085768}{945}\gamma+\frac{103352}{945}\ln(2)+\frac{75087}{70}\ln(3)\right. \nonumber\\
&&\left.
-\frac{27914012553}{67108864}\pi^4+\frac{576}{5}\zeta(3)+\frac{648724}{945}\ln(y)\right) y^9\nonumber\\
&&
-\frac{12389548}{33075}\pi y^{19/2}
+O_{\rm ln}(y^{10})\,,
\end{eqnarray}
\end{widetext}
(quadratic in $\hat a$), where we have used the dimensionless frequency variable $y$ related to $u_p$ by $u_p = y/(1-\hat ay^{3/2})^{2/3}$.

Furthermore, the GSF correction to the periastron advance for circular equatorial orbits in a Kerr spacetime in terms of the variable $y$ turns out to be
\beq
\Delta k^{{\rm circ}}=\Delta k^{{\rm circ}\,, a^0}+\hat a \Delta k^{{\rm circ}\,, a^1}+\hat a^2\Delta k^{{\rm circ}\,, a^2}\,,
\eeq
which has been already presented in Ref. \cite{Bini2019:deltakcirc}.
We will discuss below the corresponding PN expectation for completeness.

\section{PN results}

In this section we will check the first PN terms of our results by using the center-of-mass Hamiltonian description of a two-body system with spin.
Let us start by defining the spin precession frequency of the body 1 when spin couplings higher than the spin-orbit one are taken into account.  
The Hamiltonian can then be formally written as 
\begin{widetext}
\begin{eqnarray}
H({\mathbf q},{\mathbf p},{\mathbf S_1},{\mathbf S_2})&=&
H_{\rm orb}({\mathbf q},{\mathbf p})
+{\boldsymbol \Omega}_1({\mathbf q},{\mathbf p})\cdot {\mathbf S}_1
+{\boldsymbol \Omega}_2({\mathbf q},{\mathbf p})\cdot {\mathbf S}_2\nonumber\\
&+& Q^{11}{}_{jk}({\mathbf q},{\mathbf p}) {\mathbf S}_1^j{\mathbf S}_1^k 
+2Q^{12}{}_{jk}({\mathbf q},{\mathbf p}) {\mathbf S}_1^j{\mathbf S}_2^k 
+Q^{22}{}_{jk}({\mathbf q},{\mathbf p}) {\mathbf S}_2^j{\mathbf S}_2^k \nonumber\\
&+& O^{111}{}_{ijk}({\mathbf q},{\mathbf p}){\mathbf S}_1^i {\mathbf S}_1^j{\mathbf S}_1^k 
+2O^{112}{}_{ijk}({\mathbf q},{\mathbf p}) {\mathbf S}_1^i{\mathbf S}_1^j{\mathbf S}_2^k 
+2O^{122}{}_{ijk}({\mathbf q},{\mathbf p}){\mathbf S}_1^i {\mathbf S}_2^j{\mathbf S}_2^k\nonumber\\
&&+O^{222}{}_{ijk}({\mathbf q},{\mathbf p}) {\mathbf S}_2^i{\mathbf S}_2^j{\mathbf S}_2^k +O({\rm spin}^4)\,,
\end{eqnarray}
where quadrupolar and octupolar interaction terms have been included.
Here $({\mathbf q},{\mathbf p})$ are phase-space variables and $({\mathbf S_1},{\mathbf S_2})$ the spins of the two bodies. 
Omitting the explicit dependence on the variables to ease notation, the spin precession frequency follows from the spin evolution equations  (see Eqs. (3.1)--(3.4) of Ref. \cite{Damour:2007nc})
\beq
\frac{dS_1^r}{dt}=\{S_1^r,H\}\,,\qquad \{S_1^i,S_1^j\}=\epsilon^{ijk}S_1^k\equiv S_1^{ij}\,, \quad {\rm etc.}
\,.
\eeq
We find
\begin{eqnarray}
\frac{dS_1^r}{dt}&=&\Omega _{1 k}\{S_1^r,S_1^k \}
+Q^{11}{}_{jk}   \{S_1^r,S_1^j S_1^k\} 
+2Q^{12}{}_{jk}\{S_1^r, S_1^j  \}S_2^k\nonumber\\
&+& O^{111}{}_{ijk}\{S_1^r, S_1^iS_1^jS_1^k \}
+2O^{112}{}_{ijk} \{S_1^r, S_1^iS_1^j \}S_2^k
+2O^{122}{}_{ijk} \{S_1^r, S_1^i \}S_2^jS_2^k\,\nonumber\\
&=& [\Omega _{1}\times S_1]^r 
+Q^{11}{}_{jk}[ \epsilon^{rjm}S_{1m} S_1^k+  S_1^j \epsilon^{rkm}S_{1m}  ]
+2Q^{12}{}_{jk}\epsilon^{rji}S_{1i}S_2^k\nonumber\\
&+& O^{111}{}_{ijk}\{S_1^r, S_1^iS_1^jS_1^k \}
+2O^{112}{}_{ijk} \{S_1^r, S_1^iS_1^j \}S_2^k
+2O^{122}{}_{ijk} \epsilon^{rim}S_{1m}S_2^jS_2^k\,,
\end{eqnarray}
\end{widetext}
which can be cast in the form
\beq
\frac{dS_1^r}{dt}
= S_1^{rj} \Omega_{S_{1j}} \,,
\eeq
with
\begin{eqnarray}
\Omega_{S_{1j}}&=&\Omega _{1j} 
+2 Q^{11}{}_{jk} S_1^k 
+2Q^{12}{}_{jk} S_2^k 
+3 O^{111}{}_{ijk} S_1^iS_1^k\nonumber\\
&&
+4O^{112}{}_{ijk}  S_1^iS_2^k
+2O^{122}{}_{ijk}  S_2^iS_2^k\,.
\end{eqnarray}
If both spins are aligned with the orbital angular momentum ${\mathbf L}=L e_{z}$, i.e., ${\mathbf S}_1=S_1 e_{z}$ and ${\mathbf S}_2=S_2 e_{z}$, and in addition have constant magnitudes, then $\Omega_{S_{1j}}$ can only be directed along the $z$-axis too, i.e., $\Omega_{S_{1j}}=\Omega_{S_1}\delta^z_j$, implying 
\beq
\Omega_{S_1}=\frac{\partial H}{\partial S_1}\,.
\eeq

We will compute the so-defined spin precession frequency by using the center-of-mass ADM Hamiltonian, $H=H^{\rm ADM}$, with
\beq
\label{Hadmdef}
H^{\rm ADM}=m_1+m_2+\mu \hat H^{\rm ADM}\,,
\eeq
and
\beq
\hat H^{\rm ADM}=\hat H_{\rm orb}^{\rm ADM}+\hat H_{\rm SO}^{\rm ADM}+\hat H_{\rm SS}^{\rm ADM}+\hat H_{\rm SSS}^{\rm ADM}
\,,
\eeq
including linear, quadratic and cubic spin terms up to the present knowledge, namely next-to-next-to-leading-order (NNLO) for the linear-in-spin terms, next-to-leading-order (NLO) for the quadratic in spin terms and leading-order (LO) for the cubic in spin terms (see Ref. \cite{Schafer:2018kuf} for a recent review).
We will limit ourselves to the case of two point masses with aligned spins, orthogonal to the orbital motion.
We refer to Ref. \cite{Bini2019:deltaU} for the explicit expressions of the ADM Hamiltonian terms up to spin square.
Here we include also the LO cubic-in-spin term 
\begin{eqnarray}
\hat H_{\rm SSS}^{\rm ADM,LO}&=&  \left(-\frac{3}{ 4 }   \nu^2+\frac{1}{ 4 }  \Delta \nu+\frac{1}{ 8 }  +\frac{1}{ 8 } \Delta\right) \frac{L}{r^5} S_1^3\nonumber\\
&&
+\left(\frac{3}{ 4 }  \Delta \nu+\frac{3}{ 4 }   \nu+\frac{3}{ 4 }  \nu^2\right) \frac{L}{r^5} S_2 S_1^2
+ 1\leftrightarrow 2 \,,\nonumber\\
\end{eqnarray}
where the symbol $1\leftrightarrow 2$ stands for all the spin-dependent terms with the particle labels 1 and 2 exchanged ($S_1\leftrightarrow S_2$ and  $\Delta\leftrightarrow-\Delta$).

\subsection{Computing the gyroscope precession invariant}

With the ADM Hamiltonian written above and physical dimensions restored, we will compute the (averaged) spin frequency of the body 1 
\beq
\langle \Omega_{S_1} \rangle_t =\frac{1}{T_r} \oint \frac{\partial H}{\partial S_1} dt\,,
\eeq
where all phase-space variables (except to $S_1$) are kept as constant.
The analogous quantity to the spin-precession invariant \eqref{psi_def} is then defined by
\beq
\label{psi_def2}
\psi=\frac{\langle \Omega_{S_1} \rangle_t }{\Omega_\phi}\,.
\eeq
The periods of the radial and azimuthal motion as well as the associated frequencies follow from the definition
\begin{eqnarray}
T_r&=& \oint dt 
=\oint \left(\frac{\partial H}{\partial p_r}\right)^{-1}dr =2\int_0^\pi  \left(\frac{\partial H}{\partial p_r}\right)^{-1}\frac{dr}{d\chi}d\chi
\,, \nonumber\\
\Phi&=&\oint d\phi=\oint \frac{\partial H}{\partial L}dt = 2\int_0^\pi  \frac{\partial H}{\partial L}\left(\frac{\partial H}{\partial p_r}\right)^{-1} \frac{dr}{d\chi}d\chi\,,\nonumber\\
\end{eqnarray}
and 
\beq
\Omega_r=\frac{2\pi}{T_r}\,, \qquad
\Omega_\phi=\frac{\Phi}{T_r}\,,
\eeq
where we have introduced the new radial variable parametrization for eccentric (equatorial) orbits
\beq
r=\frac{1}{u (1+e\cos(\chi))} \,,
\eeq
with $u$ denoting the reciprocal of the semi-latus rectum and $e$ the eccentricity, which are now ADM variables. Both such quantities are coordinate-dependent and then gauge-dependent.
The latter should then be re-expressed in terms of a (convenient) pair of gauge invariant variables.
A convenient choice is 
\beq
\label{newvars}
\hat k=\frac{k}3\,,\qquad \iota =\frac{x}{\hat k}\,,
\eeq
which are simply related to the (fractional) periastron advance per radial period $k=\frac{\Phi}{2\pi}-1$ and the dimensionless azimuthal frequency $x = (M\Omega_\phi)^{2/3}$.
Computing these two quantities allows one to express $u$ and $e$ in terms of $\hat k$ and $\iota$, or equivalently $\iota$ and $x$ (see Ref. \cite{Bini2019:deltaU} for details). 

The spin-precession invariant \eqref{psi_def2} as a function of $\iota$ and $x$ then turns out to be
\beq
\psi(\iota,x)=\psi_{\rm S^0}(\iota,x)+\psi_{\rm S^1}(\iota,x)+\psi_{\rm S^2}(\iota,x)\,,
\eeq
with
\begin{widetext}
\begin{eqnarray}
\psi_{\rm S^0}(\iota,x)&=&
\left(\frac{3}{4}\Delta+\frac12\nu+\frac34\right)\frac{x}{\iota}\nonumber\\
&&
+\left\{
\left[\left(-\frac{9}{16}+\frac{3}{8}\nu\right)\Delta-2\nu+\frac{1}{4}\nu^2-\frac{9}{16}\right]\frac{1}{\iota}
+\left[\left(\frac{3}{4}\nu-\frac{9}{4}\right)\Delta+\frac{7}{8}\nu^2-\frac{9}{4}+\frac{11}{4}\nu\right]\frac{1}{\iota^2}
\right\}x^2\nonumber\\
&&
+\left\{
\left[\left(\frac{5}{32}\nu^2-\frac{75}{64}+\frac{3}{8}\nu\right)\Delta-\frac{93}{32}\nu-\frac{29}{96}\nu^2-\frac{75}{64}+\frac{5}{48}\nu^3\right]\frac{1}{\iota}\right.\nonumber\\
&&
+\left[\left(\frac{3}{4}\nu^2-\frac{103}{16}\nu+\frac{75}{64}+\frac{123}{512}\nu\pi^2\right)\Delta+\frac{75}{64}-\frac{141}{32}\nu+\frac{123}{512}\nu\pi^2-\frac{233}{48}\nu^2+\frac{7}{8}\nu^3+\frac{41}{256}\nu^2\pi^2\right]\frac{1}{\iota^2}\nonumber\\
&&\left.
+\left[\left(\frac{43}{2}\nu-\frac{615}{512}\nu\pi^2+\frac{27}{4}+\frac{15}{8}\nu^2\right)\Delta+\frac{37}{4}\nu-\frac{615}{512}\nu\pi^2-\frac{205}{256}\nu^2\pi^2+\frac{79}{3}\nu^2+\frac{5}{2}\nu^3+\frac{27}{4}\right]\frac{1}{\iota^3}
\right\}x^3\nonumber\\
&&
+O(x^4)
\,,
\end{eqnarray} 
\begin{eqnarray}
\psi_{\rm S^1}(\iota,x)&=&
\left\{
\left[-\frac{7}{12}\chi_1\nu+\left(\frac{1}{2}+\frac{1}{12}\nu\right)\chi_2\right]\Delta+\left(\frac{13}{12}\nu-\frac{1}{6}\nu^2\right)\chi_1+\left(\frac{1}{2}-\frac{1}{6}\nu^2-\frac{11}{12}\nu\right)\chi_2
\right\}\frac{x^{3/2}}{\iota^{3/2}}\nonumber\\
&&
+\bigg\{\left\{
\left[\left(-\frac{7}{16}\nu^2-\frac{7}{96}\nu\right)\chi_1+\left(\frac{15}{16}+\frac{13}{96}\nu+\frac{1}{16}\nu^2\right)\chi_2\right]\Delta\right.\nonumber\\
&&\left.
+\left(\frac{31}{24}\nu^2+\frac{133}{96}\nu-\frac{1}{8}\nu^3\right)\chi_1+\left(-\frac{1}{8}\nu^3-\frac{5}{24}\nu^2-\frac{167}{96}\nu+\frac{15}{16}\right)\chi_2
\right\}\frac{1}{\iota^{3/2}}\nonumber\\
&&
+\left\{
\left[\left(-\frac{125}{48}\nu^2-\frac{37}{96}\nu\right)\chi_1+\left(-\frac{45}{16}+\frac{283}{96}\nu+\frac{23}{48}\nu^2\right)\chi_2\right]\Delta+\left(-\frac{23}{24}\nu^3+\frac{31}{12}\nu^2-\frac{377}{96}\nu\right)\chi_1\right.\nonumber\\
&&\left.
+\left(-\frac{45}{16}-\frac{23}{24}\nu^3+\frac{823}{96}\nu-\frac{65}{12}\nu^2\right)\chi_2
\right\}\frac{1}{\iota^{5/2}}
\bigg\}x^{5/2}
+O(x^{7/2})
\,,
\end{eqnarray} 
and
\begin{eqnarray}
\psi_{\rm S^2}(\iota,x)&=&
\left\{
\left[\left(-\frac{7}{6}\nu+\frac{11}{24}\nu^2\right)\chi_1^2+\left(\frac{1}{2}\nu+\frac{1}{4}\nu^2\right)\chi_2\chi_1+\left(-\frac{5}{6}\nu-\frac{5}{24}\nu^2+\frac{5}{8}\right)\chi_2^2\right]\Delta+\left(\frac{7}{6}\nu-\frac{67}{24}\nu^2+\frac{1}{12}\nu^3\right)\chi_1^2\right.\nonumber\\
&&\left.
+\left(-\frac{25}{12}\nu+\frac{5}{24}\nu^2+\frac{5}{8}+\frac{1}{12}\nu^3\right)\chi_2^2+\left(\frac{1}{2}\nu+\frac{1}{6}\nu^3+\frac{7}{4}\nu^2\right)\chi_2\chi_1
\right\}\frac{x^2}{\iota^2}\nonumber\\
&&
+\bigg\{
\left\{
\left[\left(-\frac{17}{48}\nu-\frac{7}{16}\nu^2+\frac{15}{32}-\frac{5}{24}\nu^3\right)\chi_2^2+\left(-\frac{5}{24}\nu+\frac{11}{24}\nu^3-\frac{27}{16}\nu^2\right)\chi_1^2+\left(\frac{3}{8}\nu^2+\frac{3}{8}\nu+\frac{1}{4}\nu^3\right)\chi_2\chi_1\right]\Delta\right.\nonumber\\
&&
+\left(\frac{1}{12}\nu^4-\frac{31}{24}\nu-\frac{2}{3}\nu^2-\frac{1}{8}\nu^3+\frac{15}{32}\right)\chi_2^2+\left(\frac{5}{24}\nu+\frac{1}{12}\nu^4-\frac{25}{8}\nu^3+\frac{17}{24}\nu^2\right)\chi_1^2\nonumber\\
&&\left.
+\left(\frac{13}{12}\nu^3+\frac{3}{8}\nu+\frac{3}{2}\nu^2+\frac{1}{6}\nu^4\right)\chi_2\chi_1
\right\}\frac{1}{\iota^2}\nonumber\\
&&
+\left\{
\left[\left(\frac{73}{18}\nu-\frac{445}{72}\nu^2+\frac{149}{36}\nu^3\right)\chi_1^2+\left(\frac{13}{6}\nu^3+\frac{19}{4}\nu^2-6\nu\right)\chi_2\chi_1+\left(-\frac{71}{36}\nu^3-\frac{587}{72}\nu^2+\frac{929}{72}\nu-\frac{79}{16}\right)\chi_2^2\right]\Delta\right.\nonumber\\
&&
+\left(\frac{343}{24}\nu^2-\frac{73}{18}\nu-\frac{367}{18}\nu^3+\frac{8}{9}\nu^4\right)\chi_1^2+\left(-\frac{289}{12}\nu^2+\frac{8}{9}\nu^4+\frac{205}{9}\nu-\frac{79}{16}+\frac{41}{18}\nu^3\right)\chi_2^2\nonumber\\
&&\left.
+\left(\frac{16}{9}\nu^4-3\nu^2-6\nu+\frac{151}{9}\nu^3\right)\chi_2\chi_1
\right\}\frac{1}{\iota^3}
\bigg\}x^3+O(x^4)
\,,
\end{eqnarray} 
where we have used the spin variables $\chi_1$ and $\chi_2$ instead of $S_1$ and $S_2$ . 

The GSF contribution can be extracted by substituting the new variables $y=(m_2\Omega_\phi)^{2/3}$ and $\lambda=y/\hat k$, which are related to $x$ and $\iota$ by $x=y(1+q)^{2/3}$ and $\iota=\lambda(1+q)^{2/3}$, into the previous expressions, expanding them in power series of the mass ratio $q$ and selecting the first order terms.
One then gets the 1SF part 
\begin{eqnarray}
\psi_{{\rm 1SF,S^0}}(y,\lambda)&=& 
-\frac{y}{\lambda}+\left(-\frac{5}{4\lambda}+\frac{8}{\lambda^2}\right)y^2
+\left[-\frac{53}{16\lambda}+\left(\frac{123}{256}\pi^2-\frac{93}{8}\right)\frac{1}{\lambda^2}+\left(\frac{69}{4}-\frac{615}{256}\pi^2\right)\frac{1}{\lambda^3}\right]y^3
+O(y^4)\,,\nonumber\\
\psi_{{\rm 1SF,S^1}}(y,\lambda)&=& 
\left(-\frac{11}{6}\chi_2+\frac{1}{2}\chi_1\right)\frac{y^{3/2}}{\lambda^{3/2}}
+\left[\left(-\frac{107}{48}\chi_2+\frac{21}{16}\chi_1\right)\frac{1}{\lambda^{3/2}}+\left(\frac{823}{48}\chi_2-\frac{69}{16}\chi_1\right)\frac{1}{\lambda^{5/2}}\right]y^{5/2}
+O(y^{7/2})
\,,\nonumber\\
\psi_{{\rm 1SF,S^2}}(y,\lambda)&=& 
\left(-\frac{25}{6}\chi_2^2+\chi_2\chi_1\right)\frac{y^2}{\lambda^2}
+\left[\left(-\frac{47}{24}\chi_2^2+\frac{3}{4}\chi_2\chi_1\right)\frac{1}{\lambda^2}+\left(\frac{410}{9}\chi_2^2-12\chi_2\chi_1\right)\frac{1}{\lambda^3}\right]y^3
+O(y^4)
\,.
\end{eqnarray}
The last step consists in computing the Kerr background values for $y$ and $\lambda$, both functions of $u_p$ and $e_p$ (say, to distinguish them from the corresponding ADM quantities $u$ and $e$), and substituting them into the previous 1SF expressions. 
Setting $\chi_2=\hat a$ we find
\begin{eqnarray}
\psi_{\rm 1SF,S^0}(u_p,e_p)&=&
-u_p+\left(\frac{9}{4}+e_p^2\right)u_p^2
+\left[\frac{739}{16}-\frac{123}{64}\pi^2+\left(\frac{341}{16}-\frac{123}{256}\pi^2\right)e_p^2-\frac{1}{2}e_p^4\right]u_p^3
+O(u_p^4)
\,,\nonumber\\
\psi_{\rm 1SF,S^1}(u_p,e_p)&=&
\left(-\frac{1}{2}\hat a+\frac{1}{2}\chi_1\right)u_p^{3/2}
+\left[\left(-\frac{9}{8}\chi_1-\frac{1}{8}\hat a\right)e_p^2-\frac{41}{8}\hat a+\frac{3}{8}\chi_1\right]u_p^{5/2}
+O(u_p^{7/2})
\,,\nonumber\\
\psi_{\rm 1SF,S^2}(u_p,e_p)&=& 
-\hat a^2u_p^2
+\left[\left(-2\hat a^2+\frac{9}{4}\hat a\chi_1\right)e_p^2+\frac{15}{4}\hat a^2-\frac{7}{4}\hat a\chi_1\right]u_p^3
+O(u_p^4)
\,,
\end{eqnarray}
which coincide with the GSF results for $\Delta \psi$ of the previous section for $\chi_1=0$.
\end{widetext}

\subsection{Circular limit}

Let us discuss the circular limit of previous results.
The variables $\iota$ and $x$ are not independent in this limit.
Recalling the definition \eqref{newvars}, in order to express $\iota$ as a function of $x$ it is enough to use the relation $k_{\rm circ}(x)$ for the fractional periastron advance (see Eqs. (9a)--(9h) in Ref. \cite{Tiec:2013twa})
\begin{eqnarray}
k_{\rm circ}(x)&=&k_{\rm circ,O}(x)+k_{\rm circ,S}(x)+k_{\rm circ,SS}(x)\,,
\nonumber\\
k_{\rm circ,SS}(x)&=&k_{\rm circ, S_1S_2}(x)+k_{\rm circ,S_{1,2}^2}(x)\,,
\end{eqnarray}
where
\begin{widetext}
\begin{eqnarray}
k_{\rm circ,O}(x)&=&
3 x+\left(\frac{27}{2}-7\nu\right) x^2+\left(7\nu^2-\frac{649}{4}\nu+\frac{135}{2}+\frac{123}{32}\nu\pi^2\right) x^3
+O(x^4)\,,\nonumber\\
k_{\rm circ,S}(x)&=&
\left[
(-2+2\Delta+\nu)x^{3/2}
+\left(-\frac{17}{4}\Delta\nu-17-\nu^2+17\Delta +\frac{81}{4}\nu\right) x^{5/2}\right.\nonumber\\
&&
+\left. 
\left(-\frac{733}{12}\nu^2+\frac13\nu^3+\frac{11581}{48}\nu+\frac{11}{3}\Delta\nu^2-126-\frac{5317}{48}\Delta\nu+126\Delta \right)x^{7/2}
+O(x^{9/2})
\right]\chi_1+1\leftrightarrow 2 \,,
\nonumber\\
k_{\rm circ,S_1S_2}(x)&=&
\left[
3\nu x^2+\left(2\nu^2+45\nu\right)x^3
+O(x^4)\right]\chi_2\chi_1
\,,\nonumber\\
k_{\rm circ,S_{1,2}^2}(x)&=&
\left[
\left(\frac34-\frac32\nu-\frac34\Delta\right)x^2
+\left(6\nu^2-\frac{189}{4}\nu+\frac{67}{4}-\frac{67}{4}\Delta+\frac{55}{4}\Delta\nu\right)x^3
+O(x^4)\right]\chi_1^2+1\leftrightarrow 2\,,
\end{eqnarray}
so that 
\beq
\iota_{\rm circ}(x)=\frac{3x}{k_{\rm circ}(x)}\,.
\eeq
We then find 
\beq
\psi_{\rm circ}(x)=\psi_{\rm circ,S^0}(x)+\psi_{\rm circ,S^1}(x)+\psi_{\rm circ,S^2}(x)\,,
\eeq
where
\begin{eqnarray}
\psi_{\rm circ,S^0}(x)&=&
\left(\frac{3}{4}\Delta+\frac{1}{2}\nu+\frac{3}{4}\right)x
+\left[\left(-\frac{5}{8}\nu+\frac{9}{16}\right)\Delta+\frac{9}{16}+\frac{5}{4}\nu-\frac{1}{24}\nu^2\right]x^2\nonumber\\
&&
+\left[\left(\frac{27}{32}-\frac{39}{8}\nu+\frac{5}{32}\nu^2\right)\Delta+\frac{3}{16}\nu-\frac{1}{48}\nu^3+\frac{27}{32}-\frac{105}{32}\nu^2\right]x^3
+O(x^4)
\,,\nonumber\\
\psi_{\rm circ,S^1}(x)&=&
\left[-\frac{1}{2}\Delta\chi_2-\chi_1\nu+\left(\nu-\frac{1}{2}\right)\chi_2\right]x^{3/2}\nonumber\\
&&
+\left\{
\left[\frac{11}{6}\chi_1\nu+\left(-\frac{1}{4}-\frac{1}{12}\nu\right)\chi_2\right]\Delta+\left(-\frac{1}{3}\nu+\frac{7}{6}\nu^2\right)\chi_1+\left(-\frac{1}{4}+\frac{1}{6}\nu^2+\frac{5}{12}\nu\right)\chi_2
\right\}x^{5/2}
+O(x^{7/2})
\,,\nonumber\\
\psi_{\rm circ,S^2}(x)&=&
\left[\left(-\chi_2\chi_1\nu-\frac{1}{4}\chi_2^2\right)\Delta-\frac{3}{2}\chi_1^2\nu^2+(\nu^2-\nu)\chi_2\chi_1+\left(\frac{1}{2}\nu-\frac{1}{4}+\frac{1}{2}\nu^2\right)\chi_2^2\right]x^3
+O(x^4)
\,.
\end{eqnarray}
The spin orbit term $\psi_{\rm circ,S^0}$ is given by Eq. (9) of Ref. \cite{Dolan:2013roa}.
The other terms agree with those computed in Ref. \cite{Bini:2018ylh} by using the EFT results of Ref. \cite{Levi:2014sba}, which allow for the inclusion of the following further term
\begin{eqnarray}
\psi_{\rm circ,S^1\,NNLO}(x)&=&
\left\{
\left[\left(-\frac{137}{36}\nu^2+\frac{19}{4}\nu\right)\chi_1+\left(\frac{143}{48}\nu-\frac{15}{16}+\frac{53}{144}\nu^2\right)\chi_2\right]\Delta+\left(-\frac{29}{8}\nu+\frac{931}{72}\nu^2-\frac{59}{72}\nu^3\right)\chi_1\right.\nonumber\\
&&\left.
+\left(-\frac{805}{144}\nu^2+\frac{233}{48}\nu-\frac{15}{16}-\frac{53}{72}\nu^3\right)\chi_2
\right\}x^{7/2}
\,.
\end{eqnarray}
The corresponding 1SF expansion then reads
\begin{eqnarray}
\psi_{\rm circ,1SF}(y)&=& 
y^2-3y^3
+(\chi_2-\chi_1)y^{3/2}+\frac{3}{2}y^{5/2}\chi_1+\left(\frac{16}{3}\chi_2+\frac{9}{8}\chi_1\right)y^{7/2}
-2\chi_1\chi_2y^3
+O(y^4)\,,
\end{eqnarray}
which agrees with Eq. (4.8) of Ref. \cite{Bini:2018ylh} for $\chi_1=0$.

Finally, the 1SF expansion of the fractional periastron advance is 
\begin{eqnarray}
k_{\rm circ,1SF,O}(y)&=& 
2y+11y^2+\left(\frac{123}{32}\pi^2-\frac{109}{4}\right)y^3
+O(y^4)
\,,\nonumber\\
k_{\rm circ,1SF,S^1}(y)&=& 
(\chi_2-3\chi_1)y^{3/2}
+\left(\frac{11}{6}\chi_2-18\chi_1\right)y^{5/2}
+\left(-\frac{243}{2}\chi_1+\frac{385}{24}\chi_2\right)y^{7/2}
+O(y^{9/2})
\,,\nonumber\\
k_{\rm circ,1SF,S^2}(y)&=& 
(-\chi_2^2+3\chi_2\chi_1)y^2
+\left(-\frac{55}{2}\chi_2^2+45\chi_2\chi_1\right)y^3
+O(y^4)
\,,\nonumber\\
\end{eqnarray}
\end{widetext}
which agrees with previous results \cite{Bini2019:deltakcirc} for $\chi_1=0$ and $\chi_2=\hat a$.

\section{Concluding remarks}

We have analytically computed the gravitational self-force correction to the gyroscope precession along slightly eccentric equatorial orbits in the Kerr spacetime, generalizing known expressions in the Schwarzschild case. 
Our results are accurate through the 9.5PN order and to second order in both eccentricity and rotation parameter.
We have also improved to the 9.5PN level the current knowledge of the spin-precession invariant for eccentric orbits in the non-rotating case and for circular orbits in the same Kerr case.
As an independent check, we have calculated the same invariant by using the current knowledge of the ADM Hamiltonian for two point masses with aligned spins. 
The full transcription of such a high-PN analytical result within other approaches, like the EOB model, will be considered elsewhere.

\end{document}